\documentclass[12pt]{article}

\usepackage{amsfonts, amsmath, amssymb, amsthm, bbm}
\usepackage{geometry}
\usepackage{multirow}
\usepackage{hyperref}
\usepackage{float}
\usepackage{authblk}

\usepackage{graphicx}

\title{\bf{{\Large A New Causal Approach to Account for Treatment Switching in Randomized Experiments under a Structural Cumulative Survival Model}}}
\author{Andrew Ying}
\author{Eric J. Tchetgen Tchetgen}
\affil{Department of Statistics, the Wharton School, University of Pennsylvania}
\date{}

\usepackage[style = authoryear, sorting = nty, giveninits, uniquename=init, maxbibnames=99]{biblatex}
\renewbibmacro{in:}{}
\renewbibmacro*{volume+number+eid}{%
  \printfield{volume}%
  \printfield{number}%
  \setunit{\addcomma\space}%
  \printfield{eid}}
\DeclareFieldFormat[article]{number}{\mkbibparens{#1}}
\DeclareNameAlias{sortname}{given-family}
\DeclareFieldFormat[article,inbook,incollection,inproceedings,patent,thesis,unpublished]{citetitle}{#1}
\DeclareFieldFormat[article,inbook,incollection,inproceedings,patent,thesis,unpublished]{title}{#1} 
\usepackage{color}
\definecolor{darkred}{RGB}{100,0,0}
\definecolor{darkgreen}{RGB}{0,100,0}
\definecolor{darkblue}{RGB}{0,0,150}

\usepackage{hyperref}
\hypersetup{colorlinks=true, linkcolor=darkred, citecolor=darkgreen, urlcolor=darkblue}
\usepackage{url}

\usepackage{mathtools}
\mathtoolsset{showonlyrefs}

\newtheorem{thm}{Theorem}
\newtheorem*{thm*}{Theorem}

\newtheorem{prp}{Proposition}

\newtheorem{assump}{Assumption}

\theoremstyle{remark}

\newtheorem{rem}{Remark}

\def\beq{\begin{equation}} 
\def\eeq{\end{equation}}
\def\beqn{\begin{eqnarray*}}
\def\eeqn{\end{eqnarray*}}
\def\Bitem{\begin{itemize}\setlength{\itemsep}{.2in}}
\def\bitem{\begin{itemize}\setlength{\itemsep}{.05in}}
\def\eitem{\end{itemize}}
\def\Benum{\begin{enumerate}\setlength{\itemsep}{.2in}}
\def\benum{\begin{enumerate}\setlength{\itemsep}{.05in}}
\def\eenum{\end{enumerate}}
\def\bmult{\begin{multline*}}
\def\emult{\end{multline*}}
\def\bcenter{\begin{center}}
\def\ecenter{\end{center}}
\def\bframe{\begin{frame}}
\def\eframe{\end{frame}}

\def\cF{\mathcal{F}}

\def\cN{\mathcal{N}}

\def\cV{\mathcal{V}}

\def\bbH{\mathbb{H}}

\def\bbM{\mathbb{M}}

\newcommand{\E}{\operatorname{\mathbb{E}}}
\renewcommand{\P}{\operatorname{\mathbb{P}}}
\newcommand{\Var}{\operatorname{Var}}
\newcommand{\Cov}{\operatorname{Cov}}

\newcommand{\cov}[1]{\operatorname{Cov}\left(#1\right)}

\def\eps{\varepsilon}

\def\1{\mathbbm{1}}

\begin{document}
\maketitle

\begin{abstract}
\textbf{Background: }Treatment switching in a randomized controlled trial is said to occur when a patient randomized to one treatment arm switches to another treatment arm during follow-up. This can occur at the point of disease progression, whereby patients in the control arm may be offered the experimental treatment. It is widely known that failure to account for treatment switching can seriously dilute the estimated effect of treatment on overall survival. In this paper, we aim to account for the potential impact of treatment switching in a re-analysis evaluating the treatment effect of Nucleoside Reverse Transcriptase Inhibitors (NRTIs) on a safety outcome (time to first severe or worse sign or symptom) in participants receiving a new antiretroviral regimen that either included or omitted NRTIs in the Optimized Treatment That Includes or Omits NRTIs (OPTIONS) trial.

\textbf{Methods: }We propose an estimator of a treatment causal effect under a structural cumulative survival model (SCSM) that leverages randomization as an instrumental variable to account for selective treatment switching. Unlike Robins' accelerated failure time model often used to address treatment switching, the proposed approach avoids the need for artificial censoring for estimation.  We establish that the proposed estimator is uniformly consistent and asymptotically Gaussian under standard regularity conditions. A consistent variance estimator is also given and a simple resampling approach provides uniform confidence bands for the causal difference comparing treatment groups over time on the cumulative intensity scale. We develop an R package named ``ivsacim'' implementing all proposed methods, freely available to download from R CRAN. We examine the finite performance of estimator via extensive simulations.

\textbf{Results:} 357 participants in the OPTIONS trial were randomly assigned at baseline to add-NRTIs or omit-NRTIs treatment group; 93\% subsequently completed a 48-week visit. 
Using the proposed methods, we found statistically significant evidence against the sharp null hypothesis of no treatment effect on the safety outcome (P value 0.034) and our SCSM estimator revealed an increased risk for a safety outcome in participants receiving a new antiretroviral regimen that included NRTIs when compared to participants receiving a regimen that omitted NRTIs. In fact, under an SCSM encoding a constant additive hazards model, we estimated a hazards difference equal to 0.0039 (95\% CI 0.0002, 0.0075) over the 48-week follow-up.

\textbf{Conclusions:} Treatment-experienced patients with HIV infection starting a new optimized regimen will experience a higher risk of severe or worse sign or symptom. Previous analyses concluded that treatment-experienced patients with HIV infection starting a new optimized regimen can safely omit NRTIs without compromising virologic efficacy. Our analysis suggests that adding NRTIs is not only unnecessary to achieve optimal outcomes but may increase the risk for a safety outcome.

\end{abstract}

{\it Keywords: treatment switching, treatment crossover, G-estimation, instrumental variable.} 


\def\spacingset#1{\renewcommand{\baselinestretch}%
{#1}\small\normalsize} \spacingset{1}

\spacingset{1.5}

\section{Background}
\subsection{Treatment Switching}
Treatment switching (also called contamination or crossover) in a randomized controlled trial (RCT) is said to occur when a patient randomized to one treatment arm changes to another arm during the course of follow-up. This can occur at the point of disease progression, where patients in the control arm are switched to the experimental treatment in the hope of improving their prognostic. 
This can happen for a variety of reasons: a clinician may find switching treatment arms might be the best option for his/her patient; it may sometimes be pre-specified in the trial protocol as part of a dynamic treatment strategy; or driven by emerging evidence (internally or externally to the specific RCT) of the active treatment’s benefit may have broken the original trial equipoise. 
For example, \textcite{demetri2006efficacy} reported the results of a randomized controlled trial into the use of Sunitinib for the treatment of advanced gastrointestinal stromal tumours in patients for whom conventional therapy (Imatinib) had failed because of resistance or intolerance. Early trial results showed a strong benefit in favour of the new treatment in terms of time to tumour progression. It then led to change to an open label protocol and made Sunitinib available to the patients in the placebo arm, which led to the vast majority of eligible patients in the placebo arm switching to receive Sunitinib. This makes interpretation of data used in subsequent analyses difficult. More examples can be found in \textcite{cuzick1997adjusting, motzer2008efficacy, morden2011assessing}.

The estimated treatment effect on overall survival can be substantially diluted after patients switch without appropriate handling \parencite{bowden2016gaining}. Treatment switching can also bias the intent-to-treat effect by contaminating the treatment arm. Simple methods include censoring patients at the time-point where switching occurs, excluding switching patients from the analysis altogether, or modeling treatment as a time-varying covariate. Although commonly used, these methods are prone to selection bias because patients who switch treatment tend to have a different overall survival prognosis than patients who remain on their originally assigned treatment.

More rigorous approaches designed to account for this selection bias are available, including structural nested model (SNM) \parencite{robins1994adjusting, yamaguchi2004adjusting}, inverse probability of censoring weights (IPCW) \parencite{robins2000correcting}, and 2-stage adjustment methods \parencite{latimer2017adjusting} among those currently considered most promising. Simulation studies have shown that these methods tend to produce more accurate estimates of the switching-adjusted estimand than simple adjustment methods or a standard intention to treat (ITT) analysis. See \textcite{jimenez2017evaluating, latimer2018assessing, sullivan2020adjusting} for a systematic review of their applications. However, their performance can be compromised when the key underlying assumption of ``no unmeasured confounding'' is violated.

\subsection{Instrumental Variable}
To accommodate unmeasured confounding, a prevailing approach resorts to "so-called" instrumental variables (IV). Instrumental variables have a long tradition in econometrics \parencite{angrist2001instrumental, wooldridge2010econometric}. They have received increasing attention in epidemiology due to a revival of Mendelian randomization studies \parencite{davey2003mendelian, katan2004apolipoprotein}.

With time-to-event data, \textcite{tchetgen2015instrumental} demonstrated the validity of two-stage estimation approaches in additive hazard models for event times when the exposure obeys a particular location shift model. Two-stage estimation approaches \parencite{li2015instrumental, ying2019two} usually require a first-stage exposure distribution model that can be unnecessarily restrictive.
Recently, \textcite{martinussen2017instrumental} leveraged IVs using G-estimation to examine time-varying effects of a point exposure on the survival function, in the presence of unmeasured confounding. By instead modeling the structural cumulative survival function and the conditional mean of the instrumental variable given covariates, they managed to handle arbitrary exposures and IVs. 
However, all aforementioned IV methods in survival analysis that accommodate unmeasured confounding can only handle point exposure and are not readily available for treatment switching, which formally entails a time-varying treatment.

The rank-preserving structural failure time model (RPSFTM) of \textcite{robins1991correcting} also offers a solution to treatment switching leveraging randomization as an IV. The RPSFTM is usually estimated by inverting a logrank test statistic. Unfortunately, logrank-test-based estimation can be inefficient \parencite{schoenfeld1981asymptotic, robins1994adjusting}. \textcite{bowden2016gaining} proposed a weighted logrank test in an attempt to recover some information. In either approach, a major complication with estimation of RPSFTM is the need for artificial censoring to address administrative censoring, which further aggravates efficiency loss and renders estimation computationally challenging \parencite{robins1991correcting, joffe2001administrative, joffe2012g, vansteelandt2014structural}.

In this paper, we accommodate treatment switching under a structural cumulative survival model (SCSM) and demonstrate estimation and inference leveraging randomization as an IV, without requiring artificial censoring.
We further develop an asymptotic framework for inference based on a proposed recursive estimator of the SCSM which effectively extends the G-estimator under a structural cumulative failure time model for point exposure to the time-varying exposure setting \parencite{martinussen2017instrumental}. We also propose inferential tools to investigate exposure effects that vary over time. All proposed estimators and tools are implemented in the freely available R package ``ivsacim'' \parencite{ying2020ivsacim}. 

We illustrate the proposed approach in an analysis that aims to evaluate the treatment effect of Nucleoside Reverse Transcriptase Inhibitors (NRTIs) on a key safety outcome (time to first severe or worse sign or symptom) in participants receiving a new antiretroviral regimen that omitted or added NRTIs in the Optimized Treatment That Includes or Omits NRTIs (OPTIONS) trial \parencite{tashima2015regimen, tashima2015hiv}, where treatment switching is present due to possible discontinuation of the NRTIs assignment.

\subsection{The OPTIONS Trial}
The OPTIONS trial was a multi-center, open-label, prospective, randomized, controlled study evaluating the benefits and risks of omitting versus adding NRTIs to a new optimized antiretroviral regimen \parencite{tashima2015regimen}. The study population consists of HIV-infected patients for whom a PI-based regimen has failed and who have triple-class experience (NNRTIs, NRTIs, and PIs) and viral resistance. Study participants were recruited from 62 outpatient medical clinics into the trial centers across the United States from March 2008 through May 2011, with follow-up through 48 weeks (31 May 2012). The study population included HIV-1–infected persons who were at least 16 years of age; had a plasma HIV RNA level of 1000 copies/mL or more; had received a PI-based antiretroviral regimen; had previously used or had evidence of resistance to NRTIs and NNRTIs; and had acceptable laboratory values, including a calculated 2 creatinine clearance of 50 mL/min/1.73 m. Persons were ineligible if they had active hepatitis B infection, were pregnant or breastfeeding, or were using prohibited medications. A key criterion for randomization was that an individualized regimen with a cPSS greater than 2.0 could be constructed using study antiretroviral medications, excluding NRTIs. A cPSS (0 [not susceptible] to 1 [susceptible]) was calculated or assigned for each drug in a potential regimen based on the participant's prior drug exposure, virus susceptibility, and tropism result \parencite{tashima2015hiv, tashima2015regimen}.

Participants were randomly assigned either to omit or to add NRTIs after choosing an optimized regimen and an NRTI regimen. Before randomization, a cPSS was calculated for each participant for 20 different optimized regimens. One or more optimized regimens with a cPSS greater than 2.0 and NRTI regimens were recommended by the study team and sent to sites for selection before randomization.


Treatment switching occurred in this trial due to potential discontinuation of NRTI assignment. This occurred when a participant in the omit-NRTIs group started any NRTI or when a participant in the add-NRTIs group failed to initiate or permanently discontinued all NRTIs (event time was the scheduled week during which the event was recorded).

Previously \textcite{tashima2015hiv} examined treatment efficacy by comparing time to virologic failure between groups. However, they handled treatment switching by re-defining their primary outcome as a composite outcome reflecting regimen failure, defined as virologic failure or change in NRTI group assignment (whichever comes first), evaluated through 48 weeks. This ad hoc approach was not necessarily unreasonable for the primary outcome as only 10 of 357 subjects switched their treatment prior to virologic failure or censoring. However, in assessing treatment safety in this trial, the primary safety outcome was defined as time to first severe or worse sign or symptom; for which the higher number 17 of 357 subjects switched treatment during follow-up. In fact, they conducted simple safety analyses by performing a stratified log-rank tests for randomized treatment. They concluded that ``time to first severe or worse sign or symptom did not significantly differ between groups (P = 0.149)''. However, their safety analysis did not formally account for treatment switching and therefore may be biased towards the null due to contamination of the treatment arms, thus failing to reveal safety concerns. We later show that correctly accounting for treatment switching using our proposed methods reveals a concerning safety signal as evidenced by a statistically significant causal effect of treatment actually taken on safety outcomes, with exposure to an add-NRTIs treatment experiencing a higher rate of the safety outcome than the exposure to an omit-NRTIs treatment.

\section{Methods}

\subsection{Notation}
Data for subjects $i = 1, ..., n$ are treated as independent and identically distributed. Define
\begin{itemize}
\item $\tilde T_i$, safety outcome corresponding to time to first severe or worse sign or symptom; 
\item $C_i$, potential censoring time;
\item $ T_i = \min(\tilde T_i, C_i)$ denotes a subject's censored event time;
\item $\Delta_i = \mathbbm{1}(\tilde T_i\leq C_i)$ denotes a subject's observed event indicator;  
\item We introduce the counting process notation. We write $N_i(t) = \mathbbm{1}(T_i \le t, \Delta_i = 1)$ as the observed counting process, $\tilde N_i(t) = \mathbbm{1}(\tilde T_i \le t)$ the counting process of time to first severe or worse sign or symptom. We also define $Y_i(t) = \mathbbm{1}(T_i \ge t)$ and $\tilde Y_i(t) = \mathbbm{1}(\tilde T_i \ge t)$ the associated at-risk processes;
\item We assume that recorded data on treatment do not change except at discrete times $\{t_1, \cdots, t_M\}$. Thus the time varying treatment $D_i(t_m) = 1$ if subject $i$ is treated or exposed at time $t_m$, 0 otherwise, write $\bar D_i(t_m) = \{D_i(t_l): 0 < l \le m\}$. For any $t_m > T_i$, $D_i(t_m)$ is not observed. We define $D_i(t_m) = 0$ for $t_m > T_i$, so that the whole treatment process is well defined for each subject even after the outcome event has occurred;
\item $Z_i$ denotes the instrumental variable corresponding to baseline randomization in the OPTIONS trial;
\item $L_i$ denotes baseline (pre-randomization) covariates which we allow for.
\end{itemize}
We may drop the subscript $i$ when there is no confusion for simplicity. We also introduce the counterfactual outcomes framework for time-varying treatments \parencite{robins1986new, robins1987graphical},
\begin{itemize}
    \item $\tilde T(\bar d(t_m), 0)$, the potential time to event had possibly contrary to fact, the subject followed the treatment regime $\bar d(t_m)$ up to time $t_m$ and and the control treatment thereafter. We make the consistency assumption that $T = T(\bar d(t_m), 0)$ with probability one for individuals with observed $\bar D(t_m) = d(t_m)$ and $D(t_{l}) = 0$, for $l > m$. We further assume that intervening on an exposure can only affect survival after the time of that exposure, in other words, the event {$\tilde T(\bar D(t_{m - 1}), 0) \geq t_m$} occurs if and only if the event {$\tilde T(\bar D(t_l), 0) \geq t_m$} also occurs for all $l \geq m$. It follows that {$\{\tilde T \geq t\}$} and {$\{\tilde T(\bar D(t_m), 0) \ge t\}$} are the same events for $t \in [t_m, t_{m + 1})$;
    \item We write $\tilde N_{\bar d(t_m), 0}(t)$ and $\tilde Y_{\bar d(t_m), 0}(t)$ as the associated potential counting process and potential at risk process;
\end{itemize}



\subsection{Model}
We assume a structural cumulative survival model,
\begin{align}
    &\frac{\P\left[\tilde T(\bar D(t_m), 0) > t|\bar D(t_m), Z, L, \tilde T \ge t_m\right]}{\P\left[\tilde T(\bar D(t_{m - 1}), 0) > t|\bar D(t_m), Z, L, \tilde T \ge t_m\right] } \\
    &= \exp\left\{-\int_{t_m}^{t \wedge t_{m + 1}} D(t_m) dB_D(s)\right\}, \label{eq:sncaim}
\end{align}
for any $t \ge t_m$. 

This model may be interpreted as encoding for individuals still at risk for the outcome at time $t_m$ with covariates, IV and treatment history $L, Z, \bar{D}(t_m)$, the ratio of survival probabilities of remaining event free at time $t \ge t_m$ upon receiving one final blip of treatment at time $t_m-1$ versus at time $t_m$. This ratio of conditional survival probabilities is modeled by the RHS of \eqref{eq:sncaim}. As NRTIs can be safely omitted without compromising efficacy \parencite{tashima2015hiv}, $dB_D(t)$ here can be interpreted  as a measure of safety risk for those who share the same history and are treated at time $t_m$ and not thereafter, compared to individuals who are not treated after time $t_{m - 1}$. A positive value of $dB_D(t)$ implies that NTRIs increase patients' safety risk. It provides direct evidence as to whether patients should  switch to the control arm for safety reasons.

\begin{rem}
\textcite{seaman2020adjusting} investigated a general class of structural cumulative survival models (SCSMs), which in our setting and notation may be defined as 
\begin{align}
    &\frac{\P\left[\tilde T(\bar D(t_m), 0) > t|\bar D(t_m), Z, L, \tilde T \ge t_m\right]}{\P\left[\tilde T(\bar D(t_{m - 1}), 0) > t|\bar D(t_m), Z, L, \tilde T \ge t_m\right]} \\
    &= \exp\left\{-D(t_m)\gamma_m(t;\bar D(t_{m - 1}), L,Z)\right\}. 
\end{align}
This SCSM accommodates time varying effects as a function of both the survival time $t$ under consideration and the time of a final treatment blip $t_m \le t$; as well as effect heterogeneity as a function of a patient's observed history $(\bar D(t_{m - 1}), L, Z)$. Model \eqref{eq:sncaim} is a special case of the model in the above display which posits (i) that the causal effect of a final blip of treatment at time $t_{m}$ is short-lived in the sense that $\gamma_m(t;\bar D(t_{m - 1}), L,Z)=\gamma_m(t_{m+1};\bar D(t_{m - 1}), L,Z)$ for all $t \ge t_{m+1}$; and (ii) does not depend on a patient's history, i.e. $\gamma_m(t;\bar D(t_{m - 1}), L,Z)$ is a constant function of $(\bar D(t_{m - 1}), L, Z)$. Together (i) and (ii) imply:
\begin{equation}
    \gamma_m(t; L) = B_{D}(t \wedge t_{m + 1}) - B_{D}(t_m).
\end{equation}
Our simpler model specification delivers a convenient summary of the cumulative treatment effect comparing the marginal survival function under the always-treated treatment regime (i.e. the survival curve for $\tilde T(\bar 1)$) versus that of the never-treated treatment regime (i.e. the survival curve for $\tilde T(\bar 0)$) in terms of the function $B_D(t)$, our primary causal effect of interest.
That is, under model \eqref{eq:sncaim}, and the additional condition of no-current treatment value interaction \parencite{robins1994adjusting}, 
\begin{align}
    \frac{\P\left[\tilde T(\bar D(t_{m-1}),d(t_m), 0) > t|\bar D(t_{m-1}),d(t_m), Z, L, \tilde T \ge t_m\right]}{\P\left[\tilde T(\bar D(t_{m - 1}), 0) > t|\bar D(t_{m-1}),d(t_m), Z, L, \tilde T \ge t_m\right] } \\
    =\frac{\P\left[\tilde T(\bar D(t_{m-1}),d(t_m), 0) > t|\bar D(t_{m-1}),D(t_m)=0, Z, L, \tilde T \ge t_m\right]}{\P\left[\tilde T(\bar D(t_{m - 1}), 0) > t|\bar D(t_{m-1}),D(t_m)=0, Z, L, \tilde T \ge t_m\right] }, 
\end{align}   
one can readily establish that 
\begin{align}
    \frac{\P\left[\tilde T(\bar 1) > t\right]}{\P\left[\tilde T(\bar 0) > t\right]} 
    = \exp\left\{-\int_{0}^{t} 1 dB_D(s)\right\}= \exp\left\{-B_D(t)\right\}.
\end{align}
Therefore our estimand $B_D(t)$ can be interpreted as the difference in the log-marginal cumulative intensity function comparing always-treated versus never-treated regimes up to time $t$. The no-current treatment value interaction assumption essentially states that the instantaneous causal effect of one final blip of treatment at time $t_m$ among individuals who were treated at time $t_m$ is equal to that among individuals who were not treated at time $t_m$ conditional on past history.  

In comparison, \textcite{seaman2020adjusting} were interested in the more ambitious goal of learning about treatment effects under all possible treatment regimes under a sequential ignorability condition, i.e., that the time-varying treatment mechanism is not subject to unmeasured confounding. This more ambitious goal may not be identified in the OPTIONS study for two reasons: first, unlike \textcite{seaman2020adjusting}, we do not make an assumption that treatment switching is unconfounded, thus allowing for both time-fixed and time-varying unmeasured confounding of the treatment-outcome relationship, thus making identification of complex treatment effects far more challenging; second, as patients who switched treatment arms did so at most once in the OPTIONS study, investigating the causal effect of more complex treatment regimes may not be supported by the observed sample.
It is worth noting that other possible SCSM variants may be specified, including one which postulates a constant effect for each dose of treatment over time, while allowing for the magnitude of the treatment effect to depend on the timing of the final blip of treatment:
\begin{equation}
    \gamma_m(t; L) = \beta_{D, m}(t - t_m).
\end{equation}
A special case of this model may further impose that the treatment effects are constant as a function of the timing of the final treatment blip:
\begin{equation}\label{eq:const}
    \gamma_m(t; L) = \beta_{D}(t - t_m).
\end{equation}
\end{rem}


\subsection{Assumptions}
Our proposed identification strategy leverages the randomization process as an instrumental variable satisfying three key standard IV assumptions:
\begin{assump}[IV relevance]\label{assump:ivrelevance}
The instrument is associated with the exposure at $t_m$ for individuals still  at risk for the event time for all $t_m$; specifically, 
\begin{equation}
Z \not\perp D(t_m)~|~\tilde T \ge t_m, \bar D(t_{m - 1}), L,
\end{equation}
for all $t_m$.
\end{assump}
IV relevance requires that for subjects who remain at risk for the outcome event at time $t_m$, the instrument remains predictive of current treatment status even after conditioning on treatment and covariate history. This is a reasonable assumption in OPTIONS, given that individuals randomized to the active arm are more likely than the control arm to be treated over time, even upon conditioning on their history.
\begin{assump}[IV independence]\label{assump:ivindependence}
The instrument is independent of the potential outcome under no treatment, conditional on baseline covariates ,
\begin{equation}
Z \perp \tilde T(0) ~|~ L.
\end{equation}
\end{assump}
IV independence ensures that the IV itself is unconfounded (conditional on $L$). This assumption is clearly satisfied in the OPTIONS trial as $Z$ is randomized treatment assignment.
\begin{assump}[Exclusion restriction]\label{assup:ivexlusion}
The instrument has no direct causal effect on  the outcome other than through exposure, namely,
\begin{equation}
\tilde T(\bar d, z) = \tilde T(\bar d, z'),
\end{equation}
for all values of $\bar d$, $z$ and $z'$, where $\tilde T(\bar d, z)$ denotes the potential outcome had one intervened to set $Z$ and $\bar D = \{D_{t_1}, \cdots, D_{t_{M}}\}$ to $z$ and $\bar d$, respectively.
\end{assump}
The exclusion restriction rules out the possibility that randomization itself can impact safety via a pathway not involving treatment actually taken, a reasonable assumption in the OPTIONS trial to the extent that participants did not modify their behavior as a result of randomization in a manner that might influence their safety through a pathway independent of the treatment regime ultimately followed throughout the trial. There is no a priori reason to suspect that randomized assignment to the intervention adding NRTIs to a treatment regime would result in violation of this underlying assumption. 

In order to facilitate the exposition, we make a standard conditional independent censoring assumption.
\begin{assump}[Conditional independent censoring]\label{assup:condindcensoring}
\begin{equation}
C \perp (\tilde T, \bar D(t), Z)~|~L.
\end{equation}
\end{assump}
Note that although not further pursued here, the above assumption can be relaxed substantially by only requiring that $C \perp \tilde T ~|~\bar D(t_m), Z, L, T \geq t_m$, which however would require further adjustment for dependent censoring that can be achieved by standard inverse probability censoring weighting \parencite{robins2000correcting, robins2000marginal}.

\subsection{Estimation}
The crux of our approach for estimating $B_D(t)$, $t > 0$ is that once the exposure effect has been eliminated from the event time under the assumed SCSM, the ``residualized outcome'' would mimic $T(0)$ distributionally, and therefore should in principle satisfy Assumption 2. Building on this intuition, we formally establish that the parameter of interest $B_D(t)$ is in fact a solution to the following unbiased estimating equation:
\begin{equation}\label{eq:popuest}
    \E\left\{Z^c\underbrace{\exp\left(\int_0^{t-} D(s)dB_D(s)\right)}_{\text{Residualizes the at risk process}}\left[dN(t) - \underbrace{Y(t)D(t)dB_D(t)}_{\text{Residualizes the counting process}}\right]\right\} = 0,
\end{equation}
where $Z^c = Z - \E(Z|L)$.
\begin{prp}\label{prp:iden}
Under Assumptions \ref{assump:ivrelevance}, \ref{assump:ivindependence}, \ref{assup:ivexlusion}, \ref{assup:condindcensoring} and the model \eqref{eq:sncaim}, the parameter of interest $B_D(t)$ in \eqref{eq:sncaim} is the unique solution to the population estimating equation \eqref{eq:popuest}, which admits the closed form
\begin{equation}\label{eq:BDidentification}
B_D(t) = \int_0^t \frac{\E \left\{Z^c\exp\left[\int_0^{s-}D(u)dB_D(u)\right]dN(s)\right\}}{\E\left\{Z^cY(s)\exp\left[\int_0^{s-}D(u)dB_D(u)\right]D(s)\right\}},
\end{equation}
provided that $\E \{Z^cY(t)\exp[\int_0^{t-}D(s)dB_D(s)]D(t)\} \neq 0$ for all $t > 0$. 
\end{prp}

We note that IV relevance is necessary but not sufficient for $\E \{Z^cY(t)\exp[\int_0^{t-}D(s)dB_D(s)]D(t)\} \neq 0$ for all $t > 0$. Nevertheless, in settings such as the OPTIONS study in which $Z$ and $D(t)$ are positively associated conditional on $\tilde T \ge t_m, \bar D(t_{m - 1}), L,$ for all $t > t_m$ and $m$, its covariance with $Y(t)D(t)\exp\left[\int_0^{s-}D(u)dB_D(u)\right]$ can be expected to be strictly positive under IV relevance. The closed form \eqref{eq:BDidentification} of $B_D(t)$ suggests a two-step estimator. Suppose that one can correctly specify a model $\E(Z|L; \theta)$ for $E(Z|L)$ indexed by the finite dimensional parameter $\theta$. Note that in the OPTIONS trial the randomization probability $E(Z|L; \theta)=Pr(Z=1)$ is known by design. Otherwise,  assume that one can readily obtain a consistent estimator $\hat \theta$ for $\theta$ that is asymptotic linear with influence function $\epsilon_{2, i}(\theta)$, namely, $n^{1/2}(\hat \theta - \theta) =n^{-1/2} \sum_{i = 1}^n\epsilon_{2, i}(\theta)+o_p(1)$. 
An empirical analog of \eqref{eq:popuest} is thus:
\begin{equation}
    \sum_{i = 1}^n Z_i^c(\hat \theta)\exp\left(\int_0^{t-} D_i(s)d\hat B_D(s)\right)\left[dN_i(t) - Y_i(t)D_i(t)d\hat B_D(t)\right] = 0.
\end{equation}
The corresponding estimator $\hat B_D(t, \hat \theta)$ has explicit recursive form:
\begin{equation}
\sum_{i: T_i \le t} \frac{\Delta_i Z_i^c(\hat \theta)\exp\left[\sum_{l: T_l < T_i, \Delta_l = 1} D_i(T_l)d\hat B_D(T_l, \hat \theta)\right]}{\sum_{j: T_j \geq T_i} Z_j^c(\hat \theta)\exp\left[\sum_{l: T_l < T_i, \Delta_l = 1} D_i(T_l)d\hat B_D(T_l, \hat \theta)\right]D_j(T_i)}.
\end{equation}
Because of its recursive structure, and the key fact that the estimator only changes values at observed event times, we may evaluate it forward in time, with initial value $\hat B_D(0, \hat \theta) = 0$. 

Under model \eqref{eq:sncaim}, assuming the treatment effect function is a sufficiently smooth function of time, one may write, $B_D(t) = \int_0^t \beta_D(s) ds$, so that in the special case of a constant instantaneous treatment effect \eqref{eq:const}, 
$\beta_D(t) = \beta_D$ one may use the estimator 
\begin{equation}
    \hat \beta_D = \int_0^\tau w(t)d\hat B_D(t),
\end{equation}
with $w(t) = \tilde w(t)/\int_0^\tau\tilde w(s)ds$, $\tilde w(t) = \sum_{i = 1}^n R_i(t)$ and $\tau$ denoting time of end of study. In the Appendix, we establish that under sufficient regularity conditions, the estimator $\hat B_D(t, \hat \theta)$ is uniformly consistent for $B_D(t)$ and $\{\sqrt{n}(\hat B_D(t, \hat \theta) - B_D(t)):t \}$ is asymptotically equivalent to a certain Gaussian process. An explicit expression of the covariance function of the limiting Gaussian process is derived in the Appendix.

\begin{thm}\label{thm:cons}Under Assumptions \ref{assump:ivrelevance}, \ref{assump:ivindependence}, \ref{assup:ivexlusion}, \ref{assup:condindcensoring}, the model \eqref{eq:sncaim}, and given the technical Assumptions \ref{assump:ivbound}, \ref{assump:denombound}, \ref{assump:uniqsolu} listed in the Appendix, the estimator $\hat B_D(t, \hat \theta)$ is uniformly consistent for $B_D(t)$ on $[0, \tau]$, namely,
\begin{equation}
\sup_{t \in [0, \tau]} \left|\hat B_D(t, \hat \theta) - B_D(t)\right| \to 0~~ \text{a.s.}.
\end{equation}
\end{thm}

\begin{thm}\label{thm:ag} Under Assumptions \ref{assump:ivrelevance}, \ref{assump:ivindependence}, \ref{assup:ivexlusion}, \ref{assup:condindcensoring}, the model \eqref{eq:sncaim}, and given the technical Assumptions \ref{assump:ivbound}, \ref{assump:denombound}, \ref{assump:uniqsolu} listed in the Appendix, the normalized process $\sqrt{n}[\hat B_D(t, \hat \theta) - B_D(t)]$ converges weakly to a zero mean Gaussian process with variance function that can be consistently estimated by 
\begin{equation}
    \frac{1}{n} \sum_{i = 1}^n \hat \epsilon_i(t, \hat \theta)^2,
\end{equation}
with $\hat \epsilon_i(t, \hat \theta)$ defined in the Appendix equation \eqref{eq:empiricalerror}.
\end{thm}
We have developed an R package named ``ivscaim'' \parencite{ying2020ivsacim} available on R CRAN. In addition to implementing inferences based on the  estimator $\hat B_D(t, \hat \theta)$, our package also provides a goodness-of-fit test for the constant causal hazards difference model \eqref{eq:const} ($H_0:~B_D(t) = \beta_Dt$ for all $t$) and as well as for the causal null hypothesis ($H_0:~B_D(t) \equiv 0$ for all t), thus effectively extending to the time-varying treatment similar test statistics for point treatment SCSMs developed by \textcite{martinussen2017instrumental}.


\subsection{Simulation}
In order to investigate the finite sample performance  of our proposed methods, we conducted a simulation study in which we generated  $B = 1000$ data sets of i.i.d data with sample size $N =800, 1600$. For simplicity, similar to the OPTIONS trial, we omit observed baseline covariates $L$, and generate a bivariate baseline variable $U_i$ which confounds the relationship between time-varying treatment and the time to event outcome:
\begin{equation}
U_i = (U_{1, i}, U_{2, i})^\top \sim \cN\left((1.5, 1.5)^\top,
\begin{pmatrix}
1/4 & -1/6\\
-1/6 &1/4
\end{pmatrix}\right).
\end{equation}
We simulate a scenario in which initial treatment assignment $D_i(0)$ is generated as an independent Bernoulli random variable with event probability $\P(D_i(0) = 1) = 0.5$, corresponding to randomized treatment assignment so that  $Z_i = D_i(0)$. We also generate a potential treatment switching time $W_i$ for each individual according to 
\begin{equation}\label{eq:timetoswitch}
\P(W_i > t|Z_i, U_i) = \exp(-0.05 \cdot t -0.1 \cdot U_{1, i} \cdot t - 0.1 \cdot Z_i),
\end{equation}
and discretize it into a grid with step size $=0.1$.  Subject $i$ experiences treatment $Z_i$ before $W_i$ and is switched to $1 - Z_i$ right after $W_i$.  Thus, $Z_i$ and $W_i$ determine a patient's entire treatment process. By \eqref{eq:timetoswitch} we also allow both directions of treatment switching, which fits our application setting. The potential time to event $\tilde T_i(\bar d)$ is then generated according to
\begin{align}
    \P(\tilde T_i(\bar d) > t|U_i, Z_i) = \exp\left(-0.25 \cdot t - 0.1 \cdot \int_0^td(s)ds - 0 \cdot Z_i \cdot t  - 0.15 \cdot U_{2, i} \cdot t\right),
\end{align}
and the observed time to event $T_i$ is thus generated via consistency. In the appendix, we confirm that under the proposed data generating mechanism, we have that 
\begin{align}
    \frac{\P(\tilde T_i(\bar D_i(t_m), 0) > t|\bar D_i(t_m), Z_i, \tilde T_i > t_m)}{\P(\tilde T_i(\bar D(t_{m - 1}), 0) > t|\bar D_i(t_m), Z_i, \tilde T_i > t_m)}= \exp\left(- 0.1 \cdot \int_{t_m}^{t \wedge t_{m + 1}}D(s)ds\right).
\end{align}
Independent censoring was then generated  with overall rate of $22\%$. Treatment switching occurred at an approximate rate of 14\%, which is chosen to fit the application setting. 

Simulation results concerning $B_D(t)$ are given in Table \ref{tab:simuresults}, where (average) bias is reported at time points $t = 1, 2, 3$ along with corresponding coverage of 95\% confidence intervals CP($\hat B_D(t)$).

Simulation results confirm that the proposed estimator $\hat B_D(t)$ has small bias both at sample size 800 and 1600 at $t = 1, 2, 3$. When $N = 800$, estimated standard errors appear to be overstated resulting in larger coverage than the nominal level. This upward bias is especially large at $t = 3$. However, estimated standard errors match Monte Carlo standard errors as sample size goes up to 1600, and overall 95\% confidence intervals attain the nominal level.

\section{Results from IV analysis of OPTIONS trial}
\begin{figure}[H]
\centering
\includegraphics[scale = 0.5]{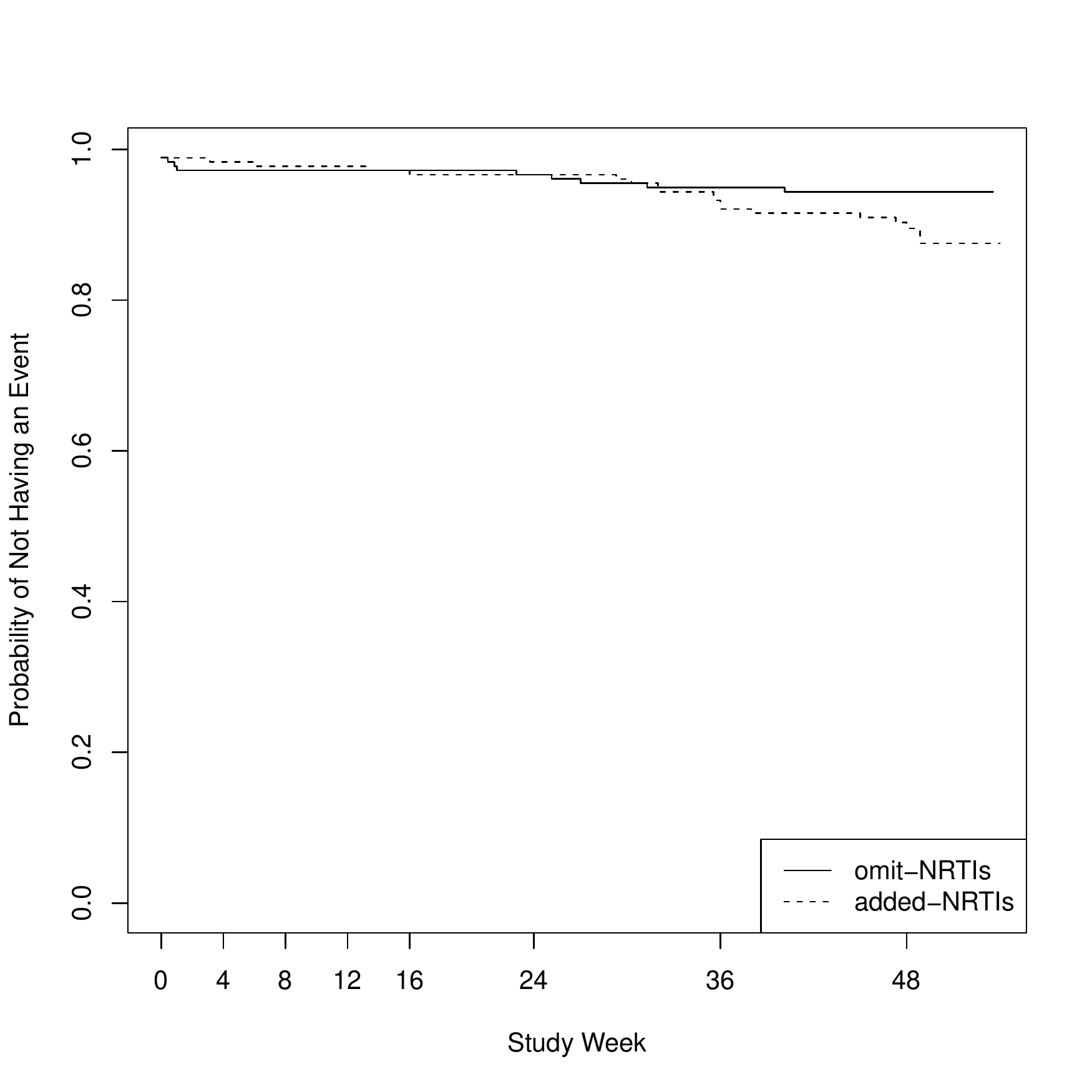}
\caption{Time to change NRTI assignment.}
\label{fig:time2insa}
\end{figure}

In the OPTIONS trial, there are 180 patients in the add-NRTIs treatment arm and 177 patients in the omit-NRTIs control arm. Among those, a total of 10 patients switched from the add-NRTIs treatment arm into the omit-NRTIs arm, while 17 patients in the omit-NRTIs group switched into the add-NRTIs arm, corresponding to  5.3\% and 9.5\% of patients switching treatment arm in the add- and omit-NRTIs groups, respectively. Of these, 9 out of 10 and 8 out of 17 treatment switching occurred before severe or worse sign or symptom in the add- and omit-NRTIs groups, respectively (Figure \ref{fig:time2insa}). 

\begin{figure}[H]
\centering
\includegraphics[scale = 0.5]{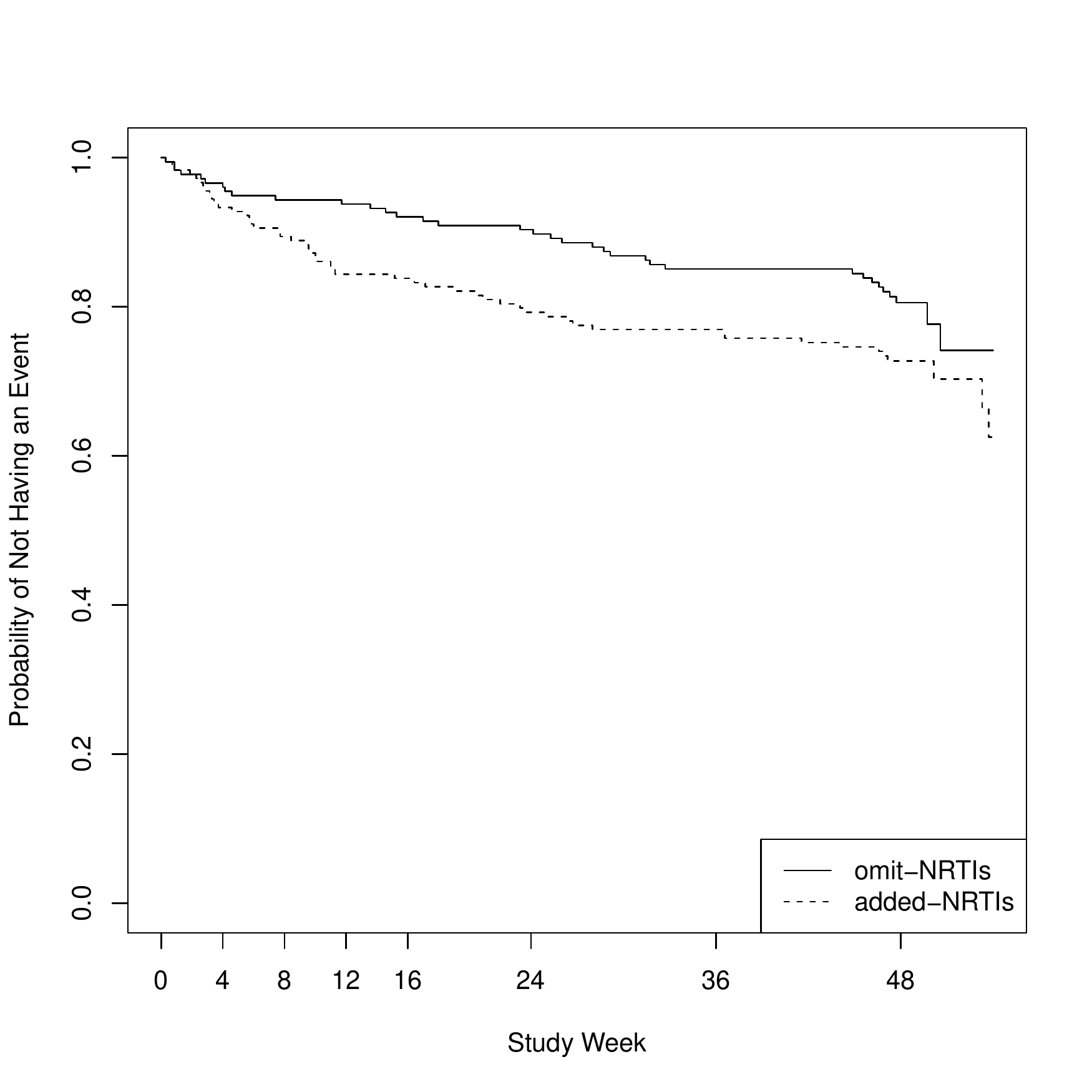}
\caption{Time to first severe or worse sign or symptom between groups.}
\label{fig:time2firstsafety}
\end{figure}

A total of 51 severe/worse sign/symptom in the add-NRTIs group and 35 in the omit-NRTIs group occurred by 48 weeks, corresponding to 28.3\% and 19.8\% in the add- and omit-NRTIs groups, respectively (Table \ref{tab:outcomesummary}). This amounted to an intent-to-treat estimated difference (i.e. a difference in outcome rates between the randomized arms), of 8.6 percentage points [95\% CI, -0.8 to 17.4 percentage points]. \textcite{tashima2015hiv} also conducted simple safety analyses by performing a stratified log-rank tests for randomized treatment and concluded that ``time to first severe or worse sign or symptom did not significantly differ between groups (P = 0.149). These served as basis for \textcite{tashima2015hiv} concluding that there was no significant statistical evidence of a safety difference between the randomized groups. However, the above analysis fails to account for treatment switching during follow-up (Figure \ref{fig:time2firstsafety}), and is therefore subject to bias towards the null hypothesis. 

\begin{figure}[H]
\centering
\includegraphics[scale = 0.5]{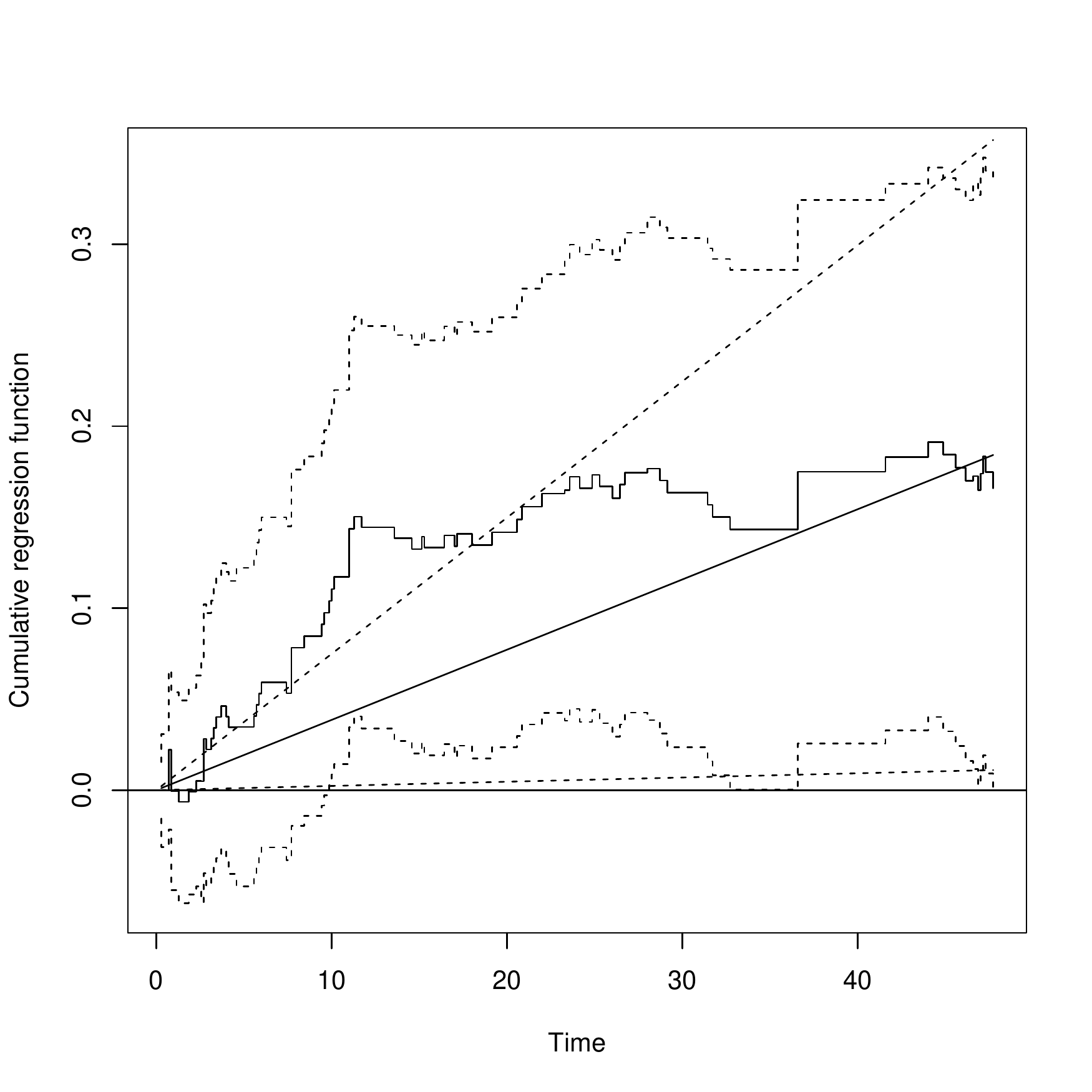}
\caption{Estimated causal effect of NRTIs on time to first severe or worse sign or symptom, along with 95\% pointwise confidence bands. The straight line corresponds to the constant effects estimator.}
\label{fig:time2firstsafetyfit}
\end{figure}

Using the proposed approach to formally account for treatment switching by leveraging randomized treatment assignment as an IV for treatment actually received which is likely confounded by unmeasured factors, we performed a test of the sharp null hypothesis of no individual causal effect, i.e.  $B_D(t) = 0$, against which we found significant statistical evidence, P value 0.034. Our approach also delivered a nonparametric estimator $\hat B_D(t)$ along with 95\% pointwise confidence bands displayed in Figure \ref{fig:time2firstsafetyfit}. From the figure, we clearly observe an excess hazard rate for experiencing the safety outcome severe/worse sign/symptom over time in the add-NRTI group compared to the omit-group. Under a constant hazards difference model, our approach estimated a hazards difference of 0.0039 (0.0002, 0.0075), P value 0.0391. Although there is significant evidence of a time-varying effect, as indicated by our goodness-of-fit test rejecting the constant effect model (P value 0.041) in favor of a time varying effect.

\section{Discussion}\label{sec:dis}
\textcite{tashima2015hiv} concluded that treatment-experienced patients with HIV infection starting a new  optimized regimen can safely omit NRTIs without compromising virologic efficacy. However, they failed to detect a significant safety issue due to a loss of power due to treatment switching. We were able to uncover a treatment difference in terms of safety which amounts to a causal effect, by leveraging randomization as an instrumental variable under a structural cumulative survival model. Our analysis suggests that adding NRTIs is not only unnecessary to achieve optimal outcomes but may in fact significantly increase the risk for a safety outcome.

Our IV analysis does not rely on unconfoundedness assumption on which standard adjustment such as inverse-probability weighting rely upon. In fact, similar to Robins g-estimation of an rank preserving structural failure time model, by leveraging randomization, our estimator can account for measured and unmeasured confounding factors driving the treatment switching mechanism. Importantly, unlike RPSFTMs, our approach obviates the need for artificial censoring for unbiasedness. In simulations studies, we found that inferences based on the proposed approach are reliable in moderate to large samples but may be somewhat conservative in small samples. We should note that as described in Section 2.2, our IV analysis does rely on three key assumptions: IV relevance, IV independence and exclusion restriction. In randomized settings such as the OPTIONS study, IV relevance and IV independence are generally expected to hold, however the exclusion restriction may be violated when, as in the OPTIONS trial, the randomized study is unblinded. This is because, one cannot rule out the possibility that patients or study physicians might modify a patient's treatment course as a result of their randomized treatment assignment in a manner that may in turn directly impact the outcome in view, thus inducing an unintended direct effect of the randomized treatment on the outcome. Such violation of the exclusion restriction can invalidate an IV analysis. We do not believe this to be a serious issue in the OPTIONS trial given that as reported in Section 3, there was no significant empirical evidence of a statistically significant association between the randomized treatment assignment on the safety outcome. Nevertheless, it would be of interest to extend the GENIUS approach of \textcite{tchetgen2017genius}, which is robust to violation of the exclusion restriction assumption, to the current setting in order to assess possible sensitivity of the proposed methodology to departure from this assumption. We plan to further investigate this possibility in future work.






\printbibliography

@article{angrist2001instrumental,
  title={Instrumental variables and the search for identification: from supply and demand to natural experiments},
  author={Angrist, Joshua D and Krueger, Alan B},
  journal={Journal of Economic Perspectives},
  volume={15},
  number={4},
  pages={69--85},
  year={2001}
}

@article{bowden2016gaining,
  title={Gaining power and precision by using model--based weights in the analysis of late stage cancer trials with substantial treatment switching},
  author={Bowden, Jack and Seaman, Shaun and Huang, Xin and White, Ian R},
  journal={Statistics in Medicine},
  volume={35},
  number={9},
  pages={1423--1440},
  year={2016},
  publisher={Wiley Online Library}
}

@article{cuzick1997adjusting,
  title={Adjusting for non-compliance and contamination in randomized clinical trials},
  author={Cuzick, Jack and Edwards, Robert and Segnan, Nereo},
  journal={Statistics in Medicine},
  volume={16},
  number={9},
  pages={1017--1029},
  year={1997},
  publisher={Wiley Online Library}
}

@article{davey2003mendelian,
  title={‘Mendelian randomization’: can genetic epidemiology contribute to understanding environmental determinants of disease?},
  author={Davey Smith, George and Ebrahim, Shah},
  journal={International Journal of Epidemiology},
  volume={32},
  number={1},
  pages={1--22},
  year={2003},
  publisher={Oxford University Press}
}

@article{demetri2006efficacy,
  title={Efficacy and safety of sunitinib in patients with advanced gastrointestinal stromal tumour after failure of imatinib: a randomised controlled trial},
  author={Demetri, George D and van Oosterom, Allan T and Garrett, Christopher R and Blackstein, Martin E and Shah, Manisha H and Verweij, Jaap and McArthur, Grant and Judson, Ian R and Heinrich, Michael C and Morgan, Jeffrey A and others},
  journal={The Lancet},
  volume={368},
  number={9544},
  pages={1329--1338},
  year={2006},
  publisher={Elsevier}
}

@article{jimenez2017evaluating,
  title={Evaluating the effects of treatment switching with randomization as an instrumental variable in a randomized controlled trial},
  author={Jimenez, Sara and Lai, Dejian and Gould, K Lance and Davis, Barry R},
  journal={Communications in Statistics-Simulation and Computation},
  volume={46},
  number={6},
  pages={4966--4990},
  year={2017},
  publisher={Taylor \& Francis}
}

@article{joffe2001administrative,
  title={Administrative and artificial censoring in censored regression models},
  author={Joffe, Marshall M},
  journal={Statistics in Medicine},
  volume={20},
  number={15},
  pages={2287--2304},
  year={2001},
  publisher={Wiley Online Library}
}

@article{joffe2012g,
  title={G-estimation and artificial censoring: problems, challenges, and applications},
  author={Joffe, Marshall M and Yang, Wei Peter and Feldman, Harold},
  journal={Biometrics},
  volume={68},
  number={1},
  pages={275--286},
  year={2012},
  publisher={Wiley Online Library}
}

@article{katan2004apolipoprotein,
  title={Apolipoprotein E isoforms, serum cholesterol, and cancer},
  author={Katan, Martjin B},
  journal={International Journal of Epidemiology},
  volume={33},
  number={1},
  pages={9--9},
  year={2004},
  publisher={Oxford University Press}
}

@article{latimer2017adjusting,
  title={Adjusting for treatment switching in randomised controlled trials--a simulation study and a simplified two-stage method},
  author={Latimer, Nicholas R and Abrams, KR and Lambert, PC and Crowther, MJ and Wailoo, AJ and Morden, JP and Akehurst, RL and Campbell, MJ},
  journal={Statistical Methods in Medical Research},
  volume={26},
  number={2},
  pages={724--751},
  year={2017},
  publisher={SAGE Publications Sage UK: London, England}
}

@article{latimer2018assessing,
  title={Assessing methods for dealing with treatment switching in clinical trials: a follow-up simulation study},
  author={Latimer, Nicholas R and Abrams, Keith R and Lambert, Paul C and Morden, James P and Crowther, Michael J},
  journal={Statistical Methods in Medical Research},
  volume={27},
  number={3},
  pages={765--784},
  year={2018},
  publisher={SAGE Publications Sage UK: London, England}
}

@article{li2015instrumental,
	Author = {Li, Jialiang and Fine, Jason and Brookhart, Alan},
	Journal = {Biometrics},
	Number = {1},
	Pages = {122--130},
	Publisher = {Wiley Online Library},
	Title = {Instrumental variable additive hazards models},
	Volume = {71},
	Year = {2015}
}

@article{lin2000semiparametric,
  title={Semiparametric regression for the mean and rate functions of recurrent events},
  author={Lin, Danyu Y and Wei, Lee-Jen and Yang, I and Ying, Zhiliang},
  journal={Journal of the Royal Statistical Society: Series B (Statistical Methodology)},
  volume={62},
  number={4},
  pages={711--730},
  year={2000},
  publisher={Wiley Online Library}
}

@article{martinussen2017instrumental,
  title={Instrumental variables estimation of exposure effects on a time-to-event endpoint using structural cumulative survival models},
  author={Martinussen, Torben and Vansteelandt, Stijn and Tchetgen Tchetgen, Eric J. and Zucker, David M},
  journal={Biometrics},
  volume={73},
  number={4},
  pages={1140--1149},
  year={2017},
  publisher={Wiley Online Library}
}

@article{morden2011assessing,
  title={Assessing methods for dealing with treatment switching in randomised controlled trials: a simulation study},
  author={Morden, James P and Lambert, Paul C and Latimer, Nicholas and Abrams, Keith R and Wailoo, Allan J},
  journal={BMC Medical Research Methodology},
  volume={11},
  number={1},
  pages={1--20},
  year={2011},
  publisher={Springer}
}

@article{motzer2008efficacy,
  title={Efficacy of everolimus in advanced renal cell carcinoma: a double-blind, randomised, placebo-controlled phase III trial},
  author={Motzer, Robert J and Escudier, Bernard and Oudard, St{\'e}phane and Hutson, Thomas E and Porta, Camillo and Bracarda, Sergio and Gr{\"u}nwald, Viktor and Thompson, John A and Figlin, Robert A and Hollaender, Norbert and others},
  journal={The Lancet},
  volume={372},
  number={9637},
  pages={449--456},
  year={2008},
  publisher={Elsevier}
}

@article{robins1986new,
  title={A new approach to causal inference in mortality studies with a sustained exposure period—application to control of the healthy worker survivor effect},
  author={Robins, James M},
  journal={Mathematical Modelling},
  volume={7},
  number={9-12},
  pages={1393--1512},
  year={1986},
  publisher={Elsevier}
}

@article{robins1987graphical,
  title={A graphical approach to the identification and estimation of causal parameters in mortality studies with sustained exposure periods},
  author={Robins, James M},
  journal={Journal of Chronic Diseases},
  volume={40},
  pages={139S--161S},
  year={1987},
  publisher={Elsevier}
}

@article{robins1991correcting,
  title={Correcting for non-compliance in randomized trials using rank preserving structural failure time models},
  author={Robins, James M and Tsiatis, Anastasios A},
  journal={Communications in Statistics-Theory and Methods},
  volume={20},
  number={8},
  pages={2609--2631},
  year={1991},
  publisher={Taylor \& Francis}
}

@article{robins1994adjusting,
  title={Adjusting for differential rates of prophylaxis therapy for PCP in high-versus low-dose AZT treatment arms in an AIDS randomized trial},
  author={Robins, James M and Greenland, Sander},
  journal={Journal of the American Statistical Association},
  volume={89},
  number={427},
  pages={737--749},
  year={1994},
  publisher={Taylor \& Francis}
}

@article{robins2000correcting,
  title={Correcting for noncompliance and dependent censoring in an AIDS clinical trial with inverse probability of censoring weighted (IPCW) log-rank tests},
  author={Robins, James M and Finkelstein, Dianne M},
  journal={Biometrics},
  volume={56},
  number={3},
  pages={779--788},
  year={2000},
  publisher={Wiley Online Library}
}

@incollection{robins2000marginal,
  title={Marginal structural models versus structural nested models as tools for causal inference},
  author={Robins, James M},
  booktitle={Statistical models in epidemiology, the environment, and clinical trials},
  pages={95--133},
  year={2000},
  publisher={Springer}
}

@article{schoenfeld1981asymptotic,
  title={The asymptotic properties of nonparametric tests for comparing survival distributions},
  author={Schoenfeld, David},
  journal={Biometrika},
  volume={68},
  number={1},
  pages={316--319},
  year={1981},
  publisher={Oxford University Press}
}

@article{seaman2020adjusting,
  title={Adjusting for time-varying confounders in survival analysis using structural nested cumulative survival time models},
  author={Seaman, Shaun and Dukes, Oliver and Keogh, Ruth and Vansteelandt, Stijn},
  journal={Biometrics},
  volume={76},
  number={2},
  pages={472--483},
  year={2020},
  publisher={Wiley Online Library}
}

@article{sullivan2020adjusting,
  title={Adjusting for Treatment Switching in Oncology Trials: A Systematic Review and Recommendations for Reporting},
  author={Sullivan, Thomas R and Latimer, Nicholas R and Gray, Jodi and Sorich, Michael J and Salter, Amy B and Karnon, Jonathan},
  journal={Value in Health},
  volume={23},
  number={3},
  pages={388--396},
  year={2020},
  publisher={Elsevier}
}

@article{tashima2015regimen,
  title={Regimen selection in the OPTIONS trial of HIV salvage therapy: drug resistance, prior therapy, and race--ethnicity determine the degree of regimen complexity},
  author={Tashima, Karen T and Mollan, Katie R and Na, Lumine and Gandhi, Rajesh T and Klingman, Karin L and Fichtenbaum, Carl J and Andrade, Adriana and Johnson, Victoria A and Eron, Joseph J and Smeaton, Laura and others},
  journal={HIV clinical trials},
  volume={16},
  number={4},
  pages={147--156},
  year={2015},
  publisher={Taylor \& Francis}
}

@article{tashima2015hiv,
  title={HIV salvage therapy does not require nucleoside reverse transcriptase inhibitors: a randomized, controlled trial},
  author={Tashima, Karen T and Smeaton, Laura M and Fichtenbaum, Carl J and Andrade, Adriana and Eron, Joseph J and Gandhi, Rajesh T and Johnson, Victoria A and Klingman, Karin L and Ritz, Justin and Hodder, Sally and others},
  journal={Annals of Internal Medicine},
  volume={163},
  number={12},
  pages={908--917},
  year={2015},
  publisher={American College of Physicians}
}

@article{tchetgen2015instrumental,
	Author = {Tchetgen Tchetgen, Eric J. and Walter, Stefan and Vansteelandt, Stijn and Martinussen, Torben and Glymour, Maria},
	Journal = {Epidemiology (Cambridge, Mass.)},
	Number = {3},
	Pages = {402-410},
	Publisher = {NIH Public Access},
	Title = {Instrumental variable estimation in a survival context},
	Volume = {26},
	Year = {2015}
}

@article{tchetgen2017genius,
  title={The GENIUS approach to robust Mendelian randomization inference},
  author={Tchetgen Tchetgen, Eric J. and Sun, BaoLuo and Walter, Stefan},
  journal={arXiv preprint arXiv:1709.07779},
  year={2017}
}

@incollection{van1996weak,
	Author = {Van Der Vaart, Aad W and Wellner, Jon A},
	Booktitle = {Weak {C}onvergence and {E}mpirical {P}rocesses},
	Pages = {16--28},
	Publisher = {Springer},
	Title = {Weak convergence},
	Year = {1996}
}

@book{van2000asymptotic,
    Author = {Van der Vaart, Aad W},
    Publisher = {Cambridge university press},
    Title = {Asymptotic {S}tatistics},
    Volume = {3},
    Year = {2000}
}

@article{vansteelandt2014structural,
  title={Structural nested models and G-estimation: the partially realized promise},
  author={Vansteelandt, Stijn and Joffe, Marshall and others},
  journal={Statistical Science},
  volume={29},
  number={4},
  pages={707--731},
  year={2014},
  publisher={Institute of Mathematical Statistics}
}

@book{wooldridge2010econometric,
  title={Econometric {A}nalysis of {C}ross {S}ection and {P}anel {D}ata},
  author={Wooldridge, Jeffrey M},
  year={2010},
  publisher={MIT press}
}

@article{yamaguchi2004adjusting,
  title={Adjusting for differential proportions of second-line treatment in cancer clinical trials. Part I: structural nested models and marginal structural models to test and estimate treatment arm effects},
  author={Yamaguchi, Takuhiro and Ohashi, Yasuo},
  journal={Statistics in Medicine},
  volume={23},
  number={13},
  pages={1991--2003},
  year={2004},
  publisher={Wiley Online Library}
}

@article{ying2019two,
  title={Two-stage residual inclusion for survival data and competing risks—An instrumental variable approach with application to SEER-Medicare linked data},
  author={Ying, Andrew and Xu, Ronghui and Murphy, James},
  journal={Statistics in Medicine},
  volume={38},
  number={10},
  pages={1775--1801},
  year={2019},
  publisher={Wiley Online Library}
}

@Manual{ying2020ivsacim,
  title = {ivsacim: Instrumental Variables Estimation Using Structural Additive Cumulative Intensity Models},
  author = {Andrew Ying},
  year = {2021},
  note = {R package version 1.1},
}
\newpage

\begin{table}[H]
\centering
\caption{Time-constant exposure effect. Bias of $\hat B_D(t)$, empirical standard error, see($\hat B_D(t)$), average estimated standard error, sd($\hat B_D(t)$),
 and coverage probability of 95\% confidence intervals CP($\hat B_D(t)$), Bias of $\hat \beta$, empirical standard error, see($\hat \beta$), average estimated standard error, sd($\hat \beta$),
 and coverage probability of 95\% pointwise confidence intervals CP($\hat \beta$), in function of sample size $n$.}
\label{tab:simuresults}
\begin{tabular}{|c|l|l|l|l|l|l|}
\hline
  Sample Sizes                     & \multicolumn{1}{l|}{}     & t = 1   & t = 2   & t = 3    & & \\ \hline
\multirow{4}{*}{n = 800} & Bias($\hat B_D(t)$)       & 0.0005 & -0.0052 & -0.0342  & Bias($\hat \beta$) &-0.0044\\
                         & see($\hat B_D(t)$)&    0.0670  & 0.1407  & 0.3064 & see($\hat \beta$) &0.1214\\ 
&sd($\hat B_D(t)$)                                & 0.0721  & 0.1492  & 0.3098 & sd($\hat \beta$) &0.1303\\ 
&95\% CP($\hat B_D(t)$)                           & 95.5  & 96.8  & 97.8 & 95\% CP($\hat \beta$) &97.6\\ \hline

 \multirow{4}{*}{n = 1600} &Bias($\hat B_D(t)$)     & -0.0021 & -0.0113 & -0.0244 & Bias($\hat \beta$) &-0.0073\\
                                &see($\hat B_D(t)$)&  0.0553  & 0.1102  & 0.2256 & see($\hat \beta$) &0.0545\\ 
                            &sd($\hat B_D(t)$)&       0.0546  & 0.1105  & 0.2228 & sd($\hat \beta$) &0.0532\\ 
                           &95\% CP($\hat B_D(t)$)&   94.7  & 95.2  & 95.7 &95\% CP($\hat \beta$) &95.7 \\ \hline
\end{tabular}
\end{table}



\begin{table}[H]
\caption{Primary outcome of virologic failure and NRTI assignment change.}
\label{tab:outcomesummary}
\resizebox{\columnwidth}{!}{%
\begin{tabular}{|l|l|l|l|}
\hline
\multirow{2}{*}{}                                & \multicolumn{2}{c|}{Patients, n (\%)}      & \multirow{2}{*}{\begin{tabular}[c]{@{}l@{}}Difference (95\% CI),\\ percentage points\end{tabular}} \\ \cline{2-3}
                                      & Add NRTIs (n = 180) & Omit NRTIs (n = 177) &                                        \\ \hline
First severe or worse sign or symptom                     & 51 (28.3)            & 35 (19.8)           & 8.6 (-0.8 to 17.9)                     \\ \hline
Change in NRTI assignment                & 10 (5.3)            & 17 (9.5)            & -4.0 (-10.1 to 2.0)                      \\ \hline
Change in NRTI assignment before failure & 9 (3.4)              & 8 (2.8)             & 0.5 (-4.4 to 5.4)                      \\ \hline
\end{tabular}
}
\end{table}


\begin{figure}[H]
\centering
\includegraphics[scale = 0.5]{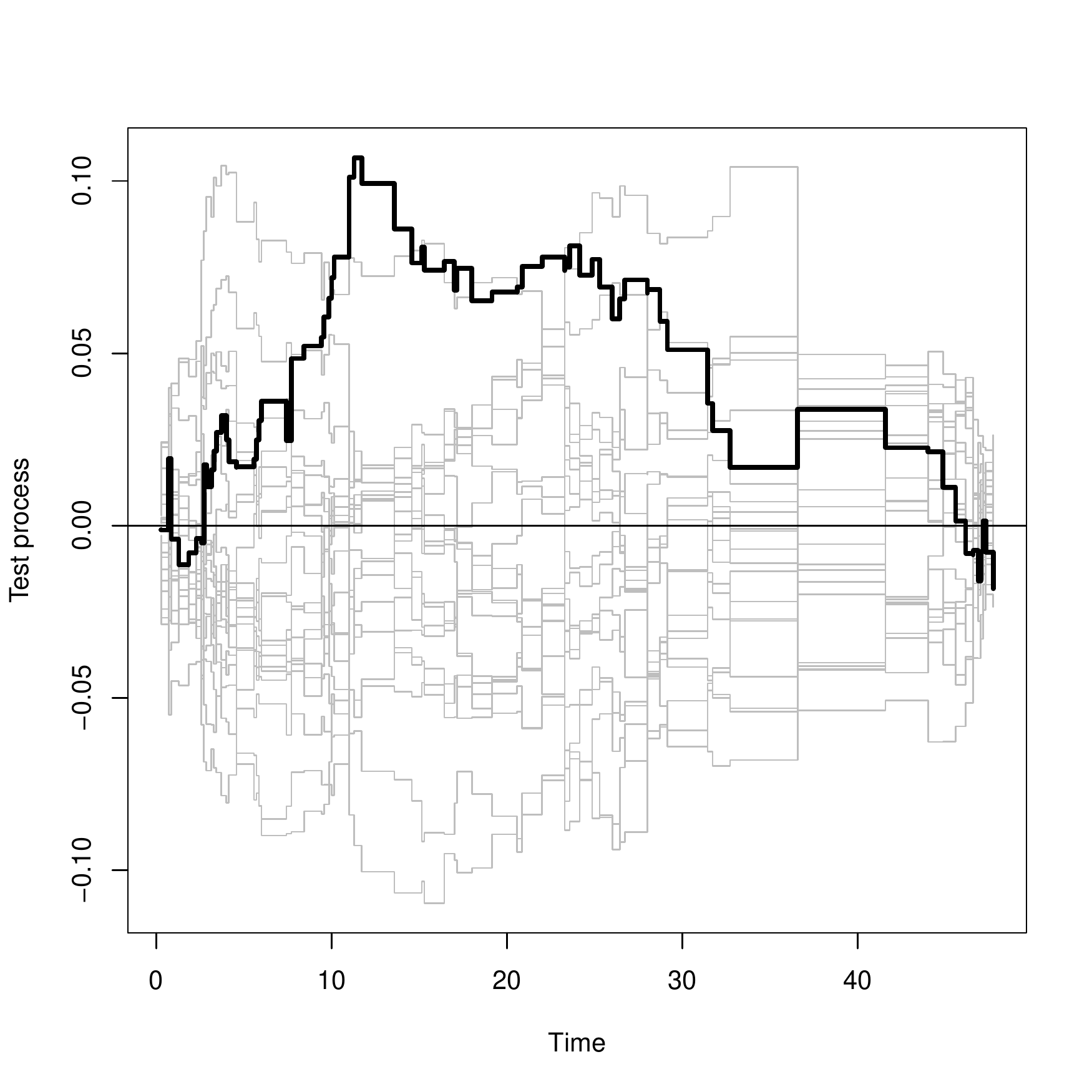}
\caption{Test process $\sqrt{n}(\hat B_D(t) - B_D(t))$ along with 20 resampled processes from its limit distribution under the null.}
\label{fig:time2firstsafetygoffit}
\end{figure}

\newpage
\appendix

\section{Proof of Proposition \ref{prp:iden}}
It is straightforward to show that \eqref{eq:sncaim} implies that 
\begin{align}
    &\E(d\tilde N_{\bar D(t_{m - 1}), 0}(t)|\bar D(t_m), Z, L, T_{\bar D(t_{m - 1}), 0} \geq t)\\
    &= \E(d\tilde N_{\bar D(t_m), 0}(t)|\bar D(t_m), Z, L, T_{\bar D(t_m), 0} \geq t) \\
    &~- D(t_m)\mathbbm{1}(t_m \leq t < t_{m + 1}) dB_D(t) \label{eq:countingdiff}.
\end{align}
For any time $t$ satisfying $t_m \le t < t_{m + 1}$, we have that under model \eqref{eq:countingdiff}) and independent censoring
\begin{eqnarray}
&&\E\bigg\{Z^ce^{\int_0^{t-} D(s)dB_D(s)}Y(t)\Big[dN(t) - D(t_m)dB_D(t)\Big]\bigg\}\\
&=&\E\bigg\{Z^ce^{\int_0^{t-} D(s)dB_D(s)}Y(t)\Big[d\tilde N_{\bar D(t_m), 0}(t) - D(t_m)dB_D(t)\Big]\bigg\}\label{eq:populationestieq}\\
&=&\E\bigg\{Z^ce^{\int_0^{t-} D(s)dB_D(s)}Y(t)\Big[\E(d\tilde N_{\bar D(t_m), 0}(t)|\bar D(t_m), Z, L, \tilde T_{\bar D(t_m), 0} \geq t) - D(t_m)dB_D(t)\Big]\bigg\}\\
&=&\E\bigg\{Z^ce^{\int_0^{t_m-} D(s)dB_D(s)}\frac{\mathbbm{1}(\tilde T(\bar D(t_m), 0) \ge t)}{e^{-\int_{t_m}^{t-} D(s)dB_D(s)}}\mathbbm{1}(C \ge t)\\
&&~~~\cdot\Big[\E(d\tilde N_{\bar D(t_m), 0}(t)|\bar D(t_m), Z, L, \tilde T_{\bar D(t_m), 0} \geq t) - D(t_m)dB_D(t)\Big]\bigg\}\\
&=&\E\bigg\{Z^ce^{\int_0^{t_m-} D(s)dB_D(s)}\P(\tilde T(\bar D(t_{m - 1}), 0) \ge t|\bar D(t_m), Z, L, \tilde T_{\bar D(t_m), 0} \geq t)\mathbbm{1}(C \ge t)\\
&&~~~\cdot\Big[\E(d\tilde N_{\bar D(t_m), 0}(t)|\bar D(t_m), Z, L, \tilde T_{\bar D(t_m), 0} \geq t) - D(t_m)dB_D(t)\Big]\bigg\}\\
&=&\E\bigg\{Z^ce^{\int_0^{t_m-} D(s)dB_D(s)}Y_{\bar D(t_{m - 1}), 0}(t)\mathbbm{1}(C \ge t)\\
&&~~~\cdot\Big[\E(d\tilde N_{\bar D(t_m), 0}(t)|\bar D(t_m), Z, L, \tilde T_{\bar D(t_m), 0} \geq t) - D(t_m)dB_D(t)\Big]\bigg\}\\
&=&\E\bigg\{Z^ce^{\int_0^{t_m-} D(s)dB_D(s)}Y_{\bar D(t_{m - 1}), 0}(t)\mathbbm{1}(C \ge t)\\
&&~~~\cdot\Big[\E(d\tilde N_{\bar D(t_{m - 1}), 0}(t)|\bar D(t_m), Z, L, \tilde T_{\bar D(t_{m - 1}), 0} \geq t)]\bigg\}\\
&=&\E\bigg\{Z^ce^{\int_0^{t_m-} D(s)dB_D(s)}Y_{\bar D(t_{m - 1}), 0}(t)\mathbbm{1}(C \ge t)d\tilde N_{\bar D(t_{m - 1}), 0}(t)\bigg\}.
\end{eqnarray}
Now we can repeat these steps in order to blip down the effect of treatment $D(t_{m - 1})$ at $t_{m - 1}$. However, note that for $t_{m - 1} < t_m \leq t$, this effect is null by assumption, and $\int_0^{t_m-}D(s)dB_D(s)$ is a function of $\bar D(t_{m - 1})$. It follows that
\begin{eqnarray}
&&\E\bigg\{Z^ce^{\int_0^{t_m-} D(s)dB_D(s)}Y_{\bar D(t_{m - 1}), 0}(t)\mathbbm{1}(C \ge t)d\tilde N_{\bar D(t_{m - 1}), 0}(t)\bigg\}\\
&&\E\bigg\{Z^ce^{\int_0^{t_m-} D(s)dB_D(s)}Y_{\bar D(t_{m - 1}), 0}(t)\mathbbm{1}(C \ge t)\E(d\tilde N_{\bar D(t_{m - 1}), 0}(t)|\bar D(t_{m - 1}), Z, L, \tilde T_{\bar D(t_{m - 1}), 0} \geq t)\bigg\}\\
&&\E\bigg\{Z^ce^{\int_0^{t_{m - 1}-} D(s)dB_D(s)}\frac{\P(T_{\bar D(t_{m - 1}), 0} \geq t|\bar D(t_{m - 1}), Z, L, \tilde T_{\bar D(t_{m - 1}), 0} \geq t)}{e^{-\int_{t_{m - 1}}^{t\wedge (t_m-)}D(s)dB_D(s)}}\mathbbm{1}(C \ge t)\\
&&~~\cdot\E(d\tilde N_{\bar D(t_{m - 1}), 0}(t)|\bar D(t_{m - 1}), Z, L, \tilde T_{\bar D(t_{m - 1}), 0} \geq t)\bigg\}\\
&&\E\bigg\{Z^ce^{\int_0^{t_{m - 1}-} D(s)dB_D(s)}\P(T_{\bar D(t_{m - 2}), 0} \geq t|\bar D(t_{m - 1}), Z, L, \tilde T_{\bar D(t_{m - 1}), 0} \geq t)\mathbbm{1}(C \ge t)\\
&&~~\cdot\E(d\tilde N_{\bar D(t_{m - 1}), 0}(t)|\bar D(t_{m - 1}), Z, L, \tilde T_{\bar D(t_{m - 1}), 0} \geq t)\bigg\}\\
&=&\E\bigg\{Z^ce^{\int_0^{t_{m - 1}-} D(s)dB_D(s)}Y_{\bar D(t_{m - 2}), 0}(t)\mathbbm{1}(C \ge t)\E(d\tilde N_{\bar D(t_{m - 1}), 0}(t)|\bar D(t_{m - 1}), Z, L, \tilde T_{\bar D(t_{m - 1}), 0} \geq t)\bigg\}\\
&=&\E\bigg\{Z^ce^{\int_0^{t_{m - 1}-} D(s)dB_D(s)}Y_{\bar D(t_{m - 2}), 0}(t)\mathbbm{1}(C \ge t)\E(d\tilde N_{\bar D(t_{m - 2}), 0}(t)|\bar D(t_{m - 1}), Z, L, \tilde T_{\bar D(t_{m - 2}), 0} \geq t)\bigg\}\\
&=&\E\bigg\{Z^ce^{\int_0^{t_{m - 1}-} D(s)dB_D(s)}Y_{\bar D(t_{m - 2}), 0}(t)\mathbbm{1}(C \ge t)d\tilde N_{\bar D(t_{m - 2}), 0}(t)\bigg\}.
\end{eqnarray} 
A recursive application of the above argument yields
\begin{eqnarray}
&&\E\bigg\{Z^c\exp\bigg(\int_0^{t-} D(s)dB_D(s)\bigg)\Big[dN_{\bar D(t_m), 0}(t) - Y(t)D(t_m)dB_D(t)\Big]\bigg\}\\
&=&\E\bigg\{Z^cY_0(t)\mathbbm{1}(C \ge t)dN_{0}(t)\bigg\} = 0,
\end{eqnarray}
by the IV independence assumption. 

Consequently, \eqref{eq:BDidentification} holds provided that $\E \{Z^cY(t)\exp[\int_0^{t-}D(s)dB_D(s)]D(t)\} \neq 0$ for all $t > 0$. 
\qed

\section{Asymptotic Results}

Throughout the paper,$\int_s^t$ is evaluated on $(s, t]$. 
The estimator can be written
\begin{equation}
\hat B_D(t, \hat \theta) = \int_0^t \frac{\sum_{i = 1}^n Z_i^c(\hat \theta)\exp\left[\int_0^{s-} D_i(u)d\hat B_D(u, \hat \theta)\right]dN_i(s)}{\sum_{j = 1}^n Z_j^c(\hat \theta)Y_j(s)\exp\left[\int_0^{s-} D_j(u)d\hat B_D(u, \hat \theta)\right]D_j(s)}.
\end{equation}

We introduce additional notation, some technical assumptions and a necessary theorem.

Let $\mu(L; \theta) = \E(Z|L; \theta)$ be the conditional mean of the instrument given observed covariates $L$, which is a function of an unknown finite-dimensional parameter $\theta$ for which we have available a consistent estimator $\hat \theta$. We assume that this estimator is also asymptotically normal with influence function $epsilon_i^\theta$ such that $\sqrt{n}(\hat \theta - \theta) = \frac{1}{\sqrt{n}}\sum_{i = 1}^n \epsilon_i^\theta + o_p(1)$.

We write $\lVert g \rVert_\infty = \sup_{t \in [0, \tau]} |g(t)|$ and use $\cV(g)$ to denote the total variation of $g$ over the interval $[0, \tau]$. Let $B^\circ(t)$ denote the true value of $B(t)$, and let $M^\circ = \lVert B^\circ \rVert_\infty < \infty$. 
\begin{assump}\label{assump:ivbound}
The instrument $Z$ is bounded by $Z_{\text{max}}$.
\end{assump}
Define $a(s, H(\cdot)) = \E[Z^cY(s)\exp(\int_0^{s-}D(u)dH(u))D(s)]$ for any $H \in \bbH$, where $\bbH$ is the set of functions on $[0, \tau]$ that have total variations bounded by some $M > M^\circ$. We make the following assumption.
\begin{assump}\label{assump:denombound}
There exists $\nu > 0$ such that $\inf_{s \in [0, \tau], H \in \bbH} a(s, H) > \nu$.
\end{assump}
Define 
\begin{equation}
    A(s, H(\cdot)) = \frac{1}{n}\sum_{i = 1}^n Z_i^cY_i(s)\exp\bigg(\int_0^{s-}D_i(u)dH(u)\bigg)D_i(s),
\end{equation}
\begin{equation}
    \Upsilon_n(H(\cdot), t) = \int_0^t \frac{\sum_{i = 1}^n Z_i^c\exp\left[\int_0^{s-}D_i(u)dH(u)\right]dN_i(s)}{nA(s, H)},
\end{equation}
\begin{equation}
    \bar \Upsilon_n(H(\cdot), t) = \int_0^t \frac{\sum_{i = 1}^n Z_i^c\exp\left[\int_0^{s-}D_i(u)dH(u)\right]dN_i(s)}{na(s, H)},
\end{equation}
and
\begin{equation}
    \Upsilon(H(\cdot), t) = \int_0^t \frac{\E\left\{Z^c\exp\left[\int_0^{s-}D(u)dH(u)\right]dN(s)\right\}}{a(s, H)}.
\end{equation}
Note that $B_n(t, \theta)$ and $B^\circ(t, \theta)$ are solutions to $B(t) = \Upsilon_n(B, t)$ and $B(t) = \Upsilon(B, t)$, respectively. We note for later reference that for any two functions $H_1$ and $H_2$, 
\begin{equation}\label{eq:gammaupperdiff}
    \lVert \Upsilon(\xi(B_1), t) - \Upsilon(\xi(B_2), t)\rVert_\infty \le 4\tau Z_{\text{max}}\exp(M)/\nu \lVert B_1 - B_2\rVert_\infty.
\end{equation}
We further assume
\begin{assump}\label{assump:uniqsolu}
$B^\circ(t)$ is the unique continuous solution to the equation $B(t) = \Upsilon(B, t)$.
\end{assump}

We shall use Helly's selection theorem to establish  Theorem \ref{thm:cons}.
\begin{thm*}[Helly's Selection Theorem]
Let $\{f_n\}$ be a sequence of functions on $[0, \tau]$ such that $\lVert f_n \rVert_\infty \le  A_1$ and $\sup_n\cV(f_n) \le A_2$, where $A_1$ and $A_2$ are finite constants. Then
\begin{enumerate}
    \item There exists a subsequence $\{f_{n_j}\}$ of $\{f_n\}$ which converges pointwise to some function $f$.
    \item If $f$ is continuous, the convergence is uniform.
\end{enumerate}
\end{thm*}

\subsection{Proof of Theorem \ref{thm:cons}}\label{sec:proofcons}

We give a roadmap for proof of consistency.
\begin{enumerate}
    \item We construct a modified version $\tilde B_n(t, \theta)$ of the estimator that is uniformly of bounded variation over $n$.
    \item We show that 
    \begin{equation}\label{eq:gammaconverge}
        \sup_{s \in [0, \tau], H \in \bbH}|\Upsilon_n(H, s) - \Upsilon(H, s)| \to 0~~ \text{a.s.}.
    \end{equation}
    \item These, together with an application of the Helly's Selection Theorem imply $\lVert\tilde B_n(t, \theta) - B^\circ(t, \theta)\rVert_\infty \to 0$ a.s..
    \item Finally we show that $\tilde B_n(t)$ is equal to $\hat B_n(t)$ in large samples, and therefore, $\sup_{t \in [0, \tau]} |\hat B_n(t, \theta) - B_D^\circ(t, \theta)| \to 0,$ a.s..
\end{enumerate}

\bigskip
\noindent
{\sc Step 1:} Let $\xi(y) = \text{sgn}(y) \min(|y|, M)$. We define the modified estimator $\tilde B_n$ to be the solution to the equation $B(t) = \Upsilon_n(\xi(B), t)$. Note that the treatment process can only take values in $0$ and $1$. Then the total variation
\begin{equation}
    \cV(\tilde B_n) \le 4\tau Z_{\text{max}}\exp(M) /\nu
\end{equation}
Therefore $\tilde B_n$ is uniformly of bounded variation over $n$. Also, since $\tilde B_n(0) = 0$, $\tilde B_n$ is uniformly bounded, therefore Helly's selection theorem applies. Note that $\Upsilon(\xi(B^\circ), t) = \Gamma(B^\circ, t) = B^\circ(t)$.

\bigskip
\noindent
{\sc Step 2:} We want to show
\begin{equation}\label{eq:barupsilonconverge}
    \sup_{s \in [0, \tau], H \in \bbH}|\bar \Upsilon_n(H, s) - \Upsilon(H, s)| \to 0, ~~\text{a.s.}.
\end{equation}
To this end, we first define
\begin{equation}
    \psi_{H,t}(T, \Delta, Z, L, D(\cdot)) := \frac{Z^c\exp(\int_0^{T-}D(s)dH(s)N(t)}{a(T, H(\cdot))},
\end{equation}
and therefore $\bar \Upsilon_n(H, t) = \P_n\psi_{H,t}$ and $\Upsilon(H, t) = \P\psi_{H,t}$. Then it suffices to show that $\{\psi_{H,t}: H \in \bbH, t \in [0, \tau]\}$ is Gilvenko-Cantelli. This result is an immediate consequence of the following facts:
\begin{enumerate}
    \item All functions that are of variation bounded form a Donsker class \cite[Example 19.11]{van2000asymptotic}. Therefore the function class $\{Z^c\exp(\int_0^{T-}D(s)dH(s))N(t): H \in \bbH,~t \in [0, \tau]\}$ is Donsker.
    \item The function $1/a(t, H)$ is a uniformly bounded, measurable function by Assumption \ref{assump:denombound}.
    \item A Donsker class multiplied by a uniformly bounded, measurable function remains Donsker \cite[Example 2.10.10]{van1996weak}.
    \item Donsker classes are Gilvenko-Cantelli.
\end{enumerate}
Similary we can show that $\{\phi_{H, s}: H \in \bbH, s \in [0, \tau]\}$ is also Gilvenko-Cantelli, where 
\begin{equation}
    \phi_{H, s}(T, \Delta, Z, L, D(\cdot)) := Z^cY(s)\exp\bigg(\int_0^{s-}D(u)dH(u)\bigg)D(s),
\end{equation}
and hence
\begin{equation}
    \sup_{s \in [0, \tau], H \in \bbH}|A(H, s) - a(H, s)| \to 0~~\text{a.s.}.
\end{equation}
This, together with Assumption \ref{assump:denombound} we get 
\begin{equation}
    \inf_{s \in [0, \tau], H \in \bbH} A(s, H) \ge \nu,
\end{equation} 
for sufficiently large $n$. We arrive at
\begin{equation}
    \sup_{s \in [0, \tau], H \in \bbH}|\bar \Upsilon_n(H, s) - \Upsilon_n(H, s)| \to 0, ~~\text{a.s.}.
\end{equation}
This together with \eqref{eq:barupsilonconverge}, implies \eqref{eq:gammaconverge}.

\bigskip
\noindent
{\sc Step 3:} We prove $\lVert\tilde B_n(t, \theta) - B(t, \theta)\rVert_\infty \to 0$ a.s. by contradiction. Without loss of generality, we assume
\begin{equation}
    \liminf_{n \to \infty}\lVert\tilde B_n(t, \theta) - B^\circ(t, \theta)\rVert_\infty > 0.
\end{equation}
By Helly's Selection Theorem, there exists a subsequence $\{n_j\}$ such that $\tilde B_{n_j}(t, \theta)$ converges to some limit $B(t)$. We further claim this limit is continuous, in fact, Lipschitz continuous. To see this, for any $t_1 < t_2$ in $[0, \tau]$, when $n_j$ is large enough
\begin{eqnarray}
|B(t_2) - B(t_1)| &\le& |B(t_2) - \tilde B_{n_j}(t_2)| + |\tilde B_{n_j}(t_2) - \tilde B_{n_j}(t_1)| + |\tilde B_{n_j}(t_1) - B(t_1)|\\
&\le& \bigg|\int_{t_1}^{t_2}\frac{\sum_{i = 1}^{n_j} Z_i^c\exp(\int_0^{s-}D_i(u)d\xi(B_{n_j})(u))dN_i(s)}{n_jA(s, \xi(B_{n_j}))}\bigg| + 2\eps\\
&\le& 4 Z_{\text{max}} \exp(M)/\nu(t_2 - t_1)+ 2\eps,
\end{eqnarray}
for any $\eps > 0$. Therefore, by the second part of Helly’s theorem, the convergence of the sub-subsequence is uniform. Going further, the limit $B$ satisfies $B(t) = \Upsilon(B, t)$ since
\begin{eqnarray}
&&\lVert B(t) -  \Upsilon(\xi(B), t)\rVert_\infty\\
&\le& \lVert B(t) -  \tilde B_{n_j}(t)\rVert_\infty + \lVert \tilde B_{n_j}(t) - \Upsilon_{n_j}(\xi(\tilde B_{n_j}(t)), t)\rVert_\infty \\
&&+ \lVert \Upsilon_{n_j}(\xi(\tilde B_{n_j}(t)), t) -  \Upsilon(B, t)\rVert_\infty + \lVert \Upsilon(\xi(\tilde B_{n_j}(t)), t) -  \Upsilon(B, t)\rVert_\infty \to 0,
\end{eqnarray}
where the last statement results from the uniform convergence of $\tilde B_{n_j}(t)$, the definition of $\tilde B_{n_j}(t)$, \eqref{eq:gammaconverge} and \eqref{eq:gammaupperdiff}. Now since the solution to $B(t) = \Upsilon(B, t)$ is unique by Assumption \ref{assump:uniqsolu} and is given by $B^\circ$, this leads to a contradiction. It hence follows that $\lVert\tilde B_n(t, \theta) - B(t, \theta)\rVert_\infty \to 0$ a.s..

\bigskip
\noindent
{\sc Step 4:} Since $\lVert B^\circ \rVert_\infty \le M^\circ$ and we just showed $\lVert \tilde B_D(t, \theta) - B_D(t, \theta)\rVert_\infty \to 0$ a.s., for sufficiently large $n$ we have $\lVert \tilde B_n \rVert_\infty \le M^\circ + \frac{1}{2}(M - M^\circ) < M$, and therefore $\xi(\tilde B_n) = \tilde B_n$. Therefore, for sufficiently large $n$, $\tilde B_n(t)$ satisfies $\tilde B_n = \Upsilon(\xi(\tilde B_n), t) =  \Upsilon(\tilde B_n, t)$, or in other words, $\hat B_n = \tilde B_n$. We thus have $\sup_{t \in [0, \tau]} |\hat B_n(t, \theta) - B^\circ(t, \theta)| \to 0,$ a.s.

The consistency of $\hat B_D (t, \hat \theta)$ then follows immediately by a Taylor series expansion since $\hat \theta$ is consistent for $\theta$.

\subsection{Proof of Theorem \ref{thm:ag}}

We aim to provide a sum of i.i.d. representation of $\sqrt{n}(\hat B(t, \hat \theta) - B(t, \theta))$ heuristically. To that end, we rewrite the normalized residual $V_n(t, \theta) := \sqrt{n}(\hat B_D(t, \theta) - B_D(t, \theta))$ at a fixed $\theta$ as a solution to a Volterra equation, which shall yield a sum of i.i.d. representation at each time point $t$ in this case. To establish this, we integrate by part the Riemann–Stieltjes integral (the integral is interpreted pathwise for the C\`adl\`ag stochastic process $D_i(s)$, $B_D$ is assumed to be continuous),
\begin{equation}\label{eq:integralbyparts}
    \int_0^{s-} D_i(u) dB_D(u) = D_i(s-)B_D(s) - \int_0^{s-} B_D(u) dD_i(u) =: G_{1, i}(s-) + G_{2, i}(s-).
\end{equation}
By Theorem \ref{thm:cons} and an application of Slutsky's Theorem, we can write 
\begin{equation}
    V_n(t, \theta) = \frac{1}{\sqrt{n}}\int_0^t \sum_{i = 1}^n H_i(s, \hat B_D) (dN_i(s) - D_i(s)dB_D(s, \theta)) + o_P(1), 
\end{equation}
where 
\begin{equation}
    H_i(s, B_D) := \frac{Z_i^c \exp(\int_0^{s-} D_i(u) d\hat B_D(u, \theta))}{\E(Z^cY(s) \exp(\int_0^{s-} D(u) dB_D(u, \theta))D(s)]}.
\end{equation}

Now we are ready to write down the Volterra equation. By consistency and a Taylor expansion, it is easy to see that
\begin{eqnarray}
V_n(t, \theta) &=& \frac{1}{\sqrt{n}}\int_0^t \sum_{i = 1}^n H_i(s, B_D)[dN_i(s) - D_i(s)dB_D(s, \theta)]\\
&&~~~+\int_0^t V_n(s-, \theta)(1 + o_P(1)) \sum_{i = 1}^n\frac{\partial H_i(s, B_D)}{\partial B_D(s-, \theta))}dN_i(s) + o_P(1), 
\end{eqnarray}
which by  \eqref{eq:integralbyparts} yields
\begin{eqnarray}
V_n(t, \theta) &=& \frac{1}{\sqrt{n}}\int_0^t \sum_{i = 1}^n H_i(s, B_D)[dN_i(s) - D_i(s)dB_D(s)]\\
&&+\int_0^t V_n(s-, \theta)(1 + o_P(1)) \sum_{i = 1}^n\frac{\partial H_i(s, B_D)}{\partial G_{1, i}(s-)}D_i(s-) dN_i(s)\\
&&+\int_0^t\sum_{i = 1}^n\frac{\partial H_i(s, B_D)}{\partial G_{2, i}(s-)}\int_0^{s-}V_n(u, \theta)(1 + o_P(1))dD_i(u) dN_i(s)+ o_P(1)\\
&=& \frac{1}{\sqrt{n}}\int_0^t \sum_{i = 1}^n H_i(s, B_D)[dN_i(s) - D_i(s)dB_D(s)]\\
&&+\int_0^t V_n(s-, \theta)(1 + o_P(1)) \sum_{i = 1}^n\frac{\partial H_i(s, B_D)}{\partial G_{1, i}(s-)}D_i(s-) dN_i(s)\\
&&+ \int_0^t V_n(s, \theta)\sum_{i = 1}^n \int_{s+}^t \frac{\partial H_i(u, B_D)}{\partial G_{2, i}(u-)} dN_i(u) dD_i(s),
\end{eqnarray}
where the last equation follows by an application of Fubini's theorem. Together we have the Volterra-equation,
\begin{eqnarray}
V_n(t, \theta) &=& \frac{1}{\sqrt{n}}\int_0^t \sum_{i = 1}^n H_i(s, B_D)[dN_i(s) - D_i(s)dB_D(s)]\\
&&+\int_0^t V_n(s-, \theta) \sum_{i = 1}^n\Big\{\frac{\partial H_i(s, B_D)}{\partial G_{1, i}(s-)} D_i(s-)dN_i(s) + \Big[\int_{s+}^t \frac{\partial H_i(u, B_D)}{\partial G_{2, i}(u-)} dN_i(u)\Big]dD_i(s)\Big\}.\label{eq:volterra}
\end{eqnarray}
which admits a solution with explicit form given by
\begin{equation}
    V_n(t, \theta) = \frac{1}{\sqrt{n}}\int_0^t \cF(s, t)\sum_{i = 1}^n H_i(s, B_D)[dN_i(s) - D_i(s)dB_D(s)] + o_P(1),
\end{equation}
where
\begin{equation}
    \cF(s, t) := \prod_{(s, t]} \bigg[1 + \sum_{i = 1}^n\Big\{\frac{\partial H_i(u, B_D)}{\partial G_{1, i}(u-)} dN_i(u) + \Big[\int_u^t\frac{\partial H_i(\cdot, B_D)}{\partial G_{2, i}(\cdot-)} dN_i(\cdot)\Big]dD_i(u)\Big\}\bigg]. 
\end{equation}
This leads to an i.i.d. representation
\begin{equation}
   V_n(t, \theta) = \frac{1}{\sqrt{n}}\sum_{i = 1}^n \epsilon_i^B(t)+ o_P(1), 
\end{equation}
with the $\epsilon_i^B(t)$’s being zero-mean i.i.d. terms. Specifically
\begin{equation}
    \epsilon_i^B(t) := \int_0^t \cF(s, t) H_i(s, B_D)[dN_i(s) - D_i(s)dB_D(s)].
\end{equation}
This, together with a Taylor expansion, gives
\begin{eqnarray}
\sqrt{n}(\hat B(t, \hat \theta) - B(t, \theta)) &=& \sqrt{n}(\hat B(t, \theta) - B(t, \theta)) + \sqrt{n}(\hat B(t, \hat \theta) - \hat B(t, \theta))\\
&=& \sqrt{n}(\hat B(t, \theta) - B(t, \theta)) + \frac{\partial \hat B(t, \theta)}{\partial \theta}\bigg|_{\hat \theta}\sqrt{n}(\hat \theta - \theta)+ o_P(1).
\end{eqnarray}
Finally we have 
\begin{equation}
    \sqrt{n}(\hat B(t, \hat \theta) - B(t, \theta)) = \frac{1}{\sqrt{n}}\sum_{i = 1}^n \epsilon_i(t, \theta) + o_P(1),
\end{equation}
where
\begin{equation}
    \epsilon_i(t, \theta) := \epsilon_i^B(t) + \frac{\partial B^\circ(t, \theta)}{\partial \theta}\bigg|_{\theta}\epsilon_i^\theta.
\end{equation}
With this i.i.d. representation, similar arguments as in \textcite{lin2000semiparametric, martinussen2017instrumental} can be adopted to establish the convergence of $V_n(t)$ in distribution to a Gaussian process.

\section{Supplementary Materials for Simulation}
We show that our data generating process is compatible with our model \eqref{eq:sncaim}, where $B_D(t) = 0.1 \cdot t$. To that end, it suffices to show that,
\begin{align}
    \frac{\P\left(\tilde T_i(\bar D_i(t_m), 0) > t|\bar D_i(t_m), Z_i, U_i, \tilde T_i \ge t_m\right)}{\P\left(\tilde T_i(\bar D_i(t_{m - 1}), 0) > t|\bar D_i(t_m), Z_i, L_i, U_i, \tilde T_i \ge t_m\right) } = \exp\left\{-0.1\int_{t_m}^{t \wedge t_{m + 1}} D(t_m) dt\right\}, \label{eq:simucondU}
\end{align}
which after integrating out $U_i$, yields \eqref{eq:sncaim}. We work on the numerator of the LHS of \eqref{eq:simucondU} first.
\begin{align}
    &\P\left(\tilde T_i(\bar D_i(t_m), 0) > t|\bar D_i(t_m), Z_i, U_i, \tilde T_i \ge t_m\right)\\
    &=\frac{\P\left(\tilde T_i(\bar D_i(t_m), 0) > t|\bar D_i(t_m), Z_i, U_i\right)}{\P\left(\tilde T_i > t_m|\bar D_i(t_m), Z_i, U_i\right)}=\frac{\P\left(\tilde T_i(\bar D_i(t_m), 0) > t|Z_i, U_i\right)}{\P\left(\tilde T_i(\bar D_i(t_m), 0) > t_m|Z_i, U_i\right)}\\
    &= \exp\left(-0.25 \cdot (t - t_m) - 0.1 \cdot \int_{t_m}^td(s)ds - 0 \cdot Z_i \cdot t  - 0.15 \cdot U_{2, i} \cdot (t - t_m)\right),
\end{align}
where the second equation follows by consistency and unconfoundedness (conditional on U) guaranteed by our data generating process. Now we switch to the denominator of the LHS of \eqref{eq:simucondU},
\begin{align}
    &\P\left(\tilde T_i(\bar D_i(t_{m - 1}), 0) > t|\bar D_i(t_m), Z_i, U_i, \tilde T_i \ge t_m\right)\\
    &=\frac{\P\left(\tilde T_i(\bar D_i(t_{m - 1}), 0) > t|\bar D_i(t_m), Z_i, U_i\right)}{\P\left(\tilde T_i \ge t_m|\bar D_i(t_m), Z_i, U_i\right)} = \frac{\P\left(\tilde T_i(\bar D_i(t_{m - 1}), 0) > t|Z_i, U_i\right)}{\P\left(\tilde T_i(\bar D_i(t_m), 0) \ge t_m|Z_i, U_i\right)}\\
    &= \exp\left(-0.25 \cdot (t - t_m) - 0 \cdot Z_i \cdot t  - 0.15 \cdot U_{2, i} \cdot (t - t_m)\right),
\end{align}
now \eqref{eq:simucondU} is straightforward.

\subsection{Variance Estimation}
A variance estimator can be obtained by plugging in empirical counterparts to unknown quantities into 
\begin{equation}\label{eq:empiricalerror}
    \hat \epsilon_i(t, \hat \theta) := \hat \epsilon_i^B(t) + \frac{\partial \hat B(t, \hat \theta)}{\partial \theta}\bigg|_{\hat \theta}\hat \epsilon_i^\theta.
\end{equation}
Note that Volterra's equation \eqref{eq:volterra} is actually a matrix equation since $\hat B_D(t)$ only changes over distinct event times $\{t_1, \cdots, t_k\}$, by writing $\hat \eps_{1, i}(t) = dN_i(t) - D_i(t)d\hat B_D(t, \hat \theta)$,
\begin{eqnarray}
&&\begin{pmatrix}
dV_n(t_1)\\
\cdots\\
dV_n(t_K)
\end{pmatrix}
=
\frac{1}{\sqrt{n}}
\begin{pmatrix}
\sum_{i = 1}^n H_i(t_1, B_D)\eps_{1, i}(t_1)\\
\cdots\\
\sum_{i = 1}^n H_i(t_K, B_D)\eps_{1, i}(t_K)\\
\end{pmatrix}
\\
&&+
\begin{pmatrix}
0 &\cdots &0 & 0\\
\sum_{i = 1}^n\frac{\partial H_i(t_1, B_D)}{\partial G_{1, i}(t_1-)} D_i(t_1-)dN_i(t_1) &\cdots &0&0\\
\cdots &\cdots &\cdots &0\\
\sum_{i = 1}^n\frac{\partial H_i(t_K, B_D)}{\partial G_{1, i}(t_K-)} D_i(t_1-)dN_i(t_K) &\cdots &\sum_{i = 1}^n\frac{\partial H_i(t_K, B_D)}{\partial G_{1, i}(t_K-)} D_i(t_{K - 1}-)dN_i(t_K) &0
\end{pmatrix}
\begin{pmatrix} 
dV_n(t_1)\\
\cdots\\
dV_n(t_K)
\end{pmatrix}.
\end{eqnarray}
That is, we have
\begin{equation}
    dV_n(t) = \eps_1 + H_ndV_n(t).
\end{equation}
The solution is therefore
\begin{equation}
    dV_n(t) = (I - H_n)^{-1}\eps_1.
\end{equation}
Note that $I - H_n$ is a triangular matrix whose inverse can by computed recursively and efficiently.


\end{document}